# On a macroscopic traversable spacewarp in practice

Mohammad Mansouryar

**Abstract:**

A design of a configuration for violation of the averaged null energy condition (ANEC) and consequently other classic energy conditions (CECs), is presented. The methods of producing effective exotic matter (EM) for a traversable wormhole (TW) are discussed. Also, the approaches of less necessity of TWs to EM are considered. The result is, TW and similar structures; i.e., warp drive (WD) and Krasnikov tube are not just theoretical subjects for teaching general relativity (GR) or objects only an advanced civilization would be able to manufacture anymore, but a quite reachable challenge for current technology. Besides, a new compound metric is introduced as a choice for testing in the lab.

## 1. Introduction:

It is clear that if a method of faster than light (FTL) travels would be discovered, its most natural consequences, such as contact to probable intelligent entities & colonizing the earth-like planets, could solve many problems of human race [1]. Even if within researching on FTL methods of communication, the related by-products could lead to rapid (not necessarily FTL) ways of transporting humans or things, that situation would be so profitable too [2].

Herein, TWs [3] are considered as the desired spacetime configurations that are able to satisfy the above requests. The main drawback of TWs hinges on their energy implications. In fact, if you want a TW, you need negative energy (NE) – energy less than vacuum energy of Minkowski 4D spacetime – with special conditions.

The stress-energy tensor of a TW metric violates both kinds of CECs [4]; pointwise & averaged. The "pointwise CECs" state, there cannot be local negative energy densities (NEDs) in physical spacetimes, while "averaged CECs" might permit violations of CECs in some points of spacetime but forbid negative values for energy measurements on the curves related to path of physical objects.

However this is not whole of the story; there are known violations of CECs, most famous of them – which would be base of present paper – is the experimentally observed Casimir effect (in parallel plate geometry with Dirichlet boundary conditions) [5].

In this paper, I have tried to review the literature, in the spirit of whether the TWs in practice are far reaching or constructible by present knowledge & technology. The conclusion is they are quite possible to manufacture provided a sufficient determination of investment on improving computation tools & necessary materials. The basic assumption is a generalization of a research done by Graham & Olum [6]. The calculations supporting my claims are too complicated, so the description will be mostly qualitative, but the principles are standard.

## 2. ANEC violation

If ANEC holds, topological censorship states: the observers that remain in the asymptotic flat region of a globally hyperbolic spacetime cannot have any experiment designed to actively detect the universe's nontrivial topology by sending & receiving causal signals (in the sense that all such signals will be homotopic [7]). With other words, it can be shown, under very general conditions, that a TW violates the ANEC in the region of the throat [8,9], by using the Raychaudhuri equation [4] together with the fact that a wormhole throat by definition defocuses light rays.

Indeed, ANEC is violated in curved spacetimes [10,11,12] – for null geodesics or non-geodesics curves – also, regarding ANEC violation, some authors [13,14] accept the existence of TWs but only in Planck scales. Further, there are other theoretical proposals [15,16,17,20], but the present approach is studying the relation between the Casimir energy – with interpretation of quantum vacuum [21] – & ANEC violation.

Qualitatively, I describe a model which can be considered as a promising generalization of the "plate with a hole" of [6].

The idea is as follows: if one considers one or several perforated pairs of mirrors (cavities), suitable for static Casimir system, in such a way that the holes are "not" symmetric, it can be expected some ANEC violation effects could leak from the extreme hole(s) of the collection. See Box 1.

For two plates with a hole in their centers they derive [6]

$$\int dx V^l V^n T_{ln} \approx 2\Delta - \frac{\boldsymbol{p}^2}{720 \ell^3} \qquad (1)$$

& declare: "above equation gives a positive result as long as $\ell > 1.6d$, which surely includes its entire range of applicability". But a modification might yield a different result. The modification is: The holes must not be (quite) face to face.

Actually, asymmetric perforated mirrors model has some advantages: (1) If the separation between the plates $\ell$ is much smaller than the radius of the hole $d$, the two plates are not equivalent to a single plate; in contrast to [6]. (2) If a null geodesics or a non-geodesics curve can pass through both of the holes, the probability of defocusing that ray – equivalent to demonstration of ANEC violation – is more than [6]. (3) Another promising feature is, necessity of concentration to Eq. (2) instead a special type of it. It means as a different point of view in p. 11 of [6], if one calculates

$$T_{ln} V^l V^n = \dot{\boldsymbol{f}}^2 + \sum_i (v_i \partial_i \boldsymbol{f})^2 \qquad (2)$$

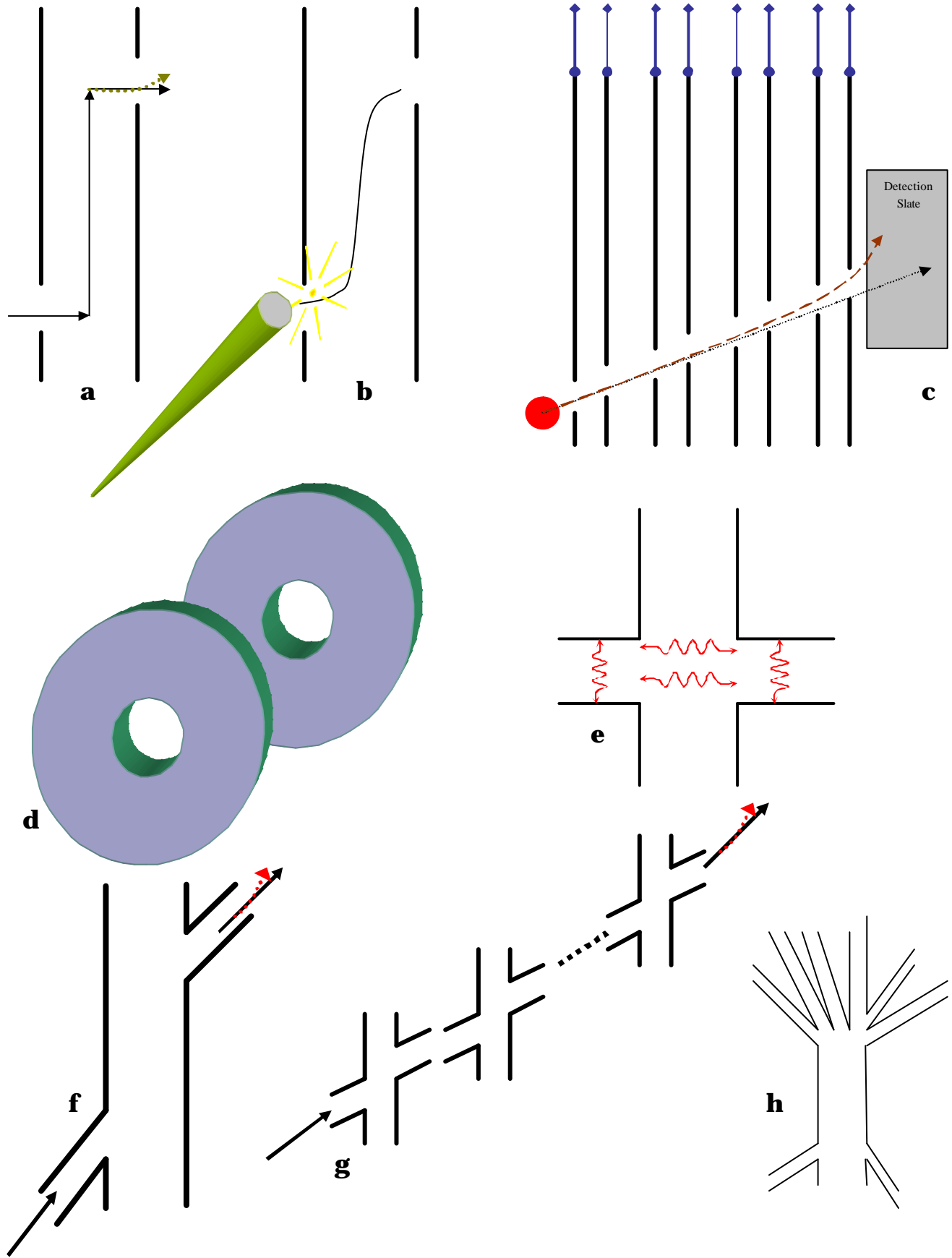

**Box of figs 1**: (**a**) The ideal path of a null ray. The ray tries to have minimum presence in the regions contained of positive contributions of the holes & positive divergences related to the walls. It quickly runs away from the entrance hole & quickly passes the region around the egress hole (i.e., along a straight line). Such a presence in the middle regions – to detect NED – is confirmed in real conditions too [23,24]. (**b**) A differentiable curve that – in contrast to situation of (**a**) but very similar – seems a natural candidate for geodesics or non-geodesic rays. However, it is reasonable there are so few of them in any group of light rays, but the scattering theory doesn't seem to forbid such a state completely. (**c**) A collection of cavities for detection of ANEC violating effects. Each plate can have more than one hole. Red circle denotes a light source which sends a ray (black arrow) to a detection slate; The blue tentacles are for engineering requirements & must not be conductive (to least interference effects). Their motions perturb the system but probably less than one already expects [27].

If ANEC violation effects occur, then there must be a defocusing of the ray (shown exaggeratedly by brown arrow, however based on AWEC violation [22], that occurs for timelike particles, at least at the end of each non-perforated cavity). In extreme conditions, it is probable the ray even doesn't arrive to the slate & impacts to a mirror; such a state can be achieved for a collection of numerous cavities. Now, some points should be noted: (*1*) More holes, more probability of detection of defocused rays, but also more positive contribution to the system, against the ANEC violation. (*2*) Less area (or radius) of the holes, less probability of detection of defocused rays, but also less positive contribution to the system in favor of ANEC violation. Therefore, an elaborate balance between the conditions (*1*) & (*2*) would be desired.

Also, there are other subtleties; if in a rather thick plate one has a circular hole (**d**), the Casimir energy would be positive there [30], that is given by (for perfect spherical boundaries):

$$E = \frac{1}{2i}\sum_{l=1}^{\infty}(2l+1)\int_{-\infty}^{\infty}\frac{dw}{2p}e^{-iwt}\int_{0}^{\infty}r^2 dr\left(2k^2[\tilde{F}_l+\tilde{G}_l](r,r)+\frac{1}{r^2}\frac{d}{dr}r\left\{\frac{d}{dr'}r'[\tilde{F}_l(r,r')+\tilde{G}_l(r,r')]\right\}_{r'=r}\right) \quad (B1)$$

However, another possibility (generally, in curved geometries), is giving negative contribution according to the curvature of spacetime (see [31, *ref 13 therein*]), that's similar for the plates themselves; although, any shape manipulations – which would require heuristic calculations – must maintain previous efforts of extracting the EM). So, much precision is required for the holes (e.g., peculiar shapes or high stretchability to yield essential geometry may be needed [111]).

Next idea is fixing smaller plates on the holes. The vertical such plates have technical troubles, & low theoretical value (**e**), because of weakening the NE contributions by concluded imposition; while oriented attached cavities have interested results (**f**). Their effect causes the ray does not feel that has got rid of a cavity (by being present in the middle regions), & chains of cavities are expecting to give negative contributions, undertake the duty of their last colleagues, & 'knead' the ray in an CEC violating manner (**g**). Besides, instead of "*one* small continuing" cavity, branches of them are imaginable (**h**), to be an element of the chains or generally any tree-structural pattern as (**g**).

*Additional remark:* There are three factors interested to a NED detection; 1) finite plate, 2) a non-complete geodesic or non-geodesic ray, 3) {in relation to Eq. (3a)}, $d$ be increased, no globally chance, so that should be increased locally, corresponding to perforated (& rather nonstationary) plates [27] (asymmetry & thoughtful dynamics are further auxiliary & definitive factors to that end). There are also similar – & probably weaker – ideas on *m* reduction (locally again). See next page.

because of being more general of "an asymmetric trajectory of a ray" rather than a ray passing in the perpendicular direction through two symmetric holes in one axis as in [6], the argument of failing the ANEC violation due to difference of a total derivative to the case of NEC contributions cannot be applied anymore & one should mention more complicated arguments for ANEC survival.

Calculations for proving that guess is too difficult, but there are facts which make us hopeful: Due to [22] for a perfectly reflecting boundary at $x = 0$, the positive energy density (PED) associated with the wall declines exponentially as $r_c(x) \sim \exp(-4|x|/a)$, while the NED associated with $f$ declines only as a power law. For large enough $x$, the NE will dominate, & the total energy will approach the form of $r_f(x) = -1/(32\pi x^3)$. That is the same for realistic situations which CECs violation is more difficult to achieve than idealized models [23,24].

On the other hand, when the ray enters or exits, the hole gives a positive contribution in favor of ANEC, but in asymmetric model the ray has more time to be in the NED regions for compensation & decreases its absorbed positive contribution. Therefore the scenario can be like this: Set up a configuration of mirrors & holes which the rays can escape from the holes & the regions near the mirrors as fast as they can, & as much as possible; thus those would spend most of their time in the middle regions which NEDs are dominant [25].

As an encouraging instance, let us consider a situation for better intuition. Visser in his book (see [3], pp. 123,124), derives an inequality in a Casimir system as a necessary condition of the "formal" ANEC violation (i.e, for "every complete" [32] null geodesic, passing through cavity) in the case of two "infinite" metallic parallel plates:

$$a < d\left(\frac{\pi^2 \hbar}{360\, mLc}\right)^{1/3} \ll d \tag{3a}$$

where $c$ is the speed of light, $a$, separation between the plates, $m$, atomic mass of the metal, $L$, thickness of the plate, & $d$, lattice spacing.

Let us test Eq. (3a) in the case of plates made of "stable metallic hydrogen" [33]. Therefore $m = m_p = 1.67 \times 10^{-27}\, Kg$, $d = 2 \times (Bohr\ radius)$, $L = Nd$, which $N$ is the number of layers [34]. Thus one has:

$$a < \frac{d^{2/3}}{\sqrt[3]{N}} 179\, nm \ll d \tag{3b}$$

Based on Eq. (3a), the mirrors should be light & hollow, however because of the fraction $(d^2/m)^{1/3}$, hollowness is more effective, its physical reason is trivial: if you could produce a metallic composite with a lot of empty spaces inside, those regions are filled by the electronic sea, & less mass of the electrons rather than any other involved particle, is in favor of being affected of the plate's whole mass by attractive force of quantum vacuum; a demand that "averaged" CECs imply.

Along with searching for materials which are not heavy (metallic hydrogen is the best!) & dense neither, one may think of perforated mirrors as "locally" hollow systems (although, that's not a standard statement, but helps to a better intuition). In that direction, if in Eq. (3b), the mirrors were about $10^4$ order of magnitudes closer (for $N=1$, & ignoring realistic nature beneath the plasma wavelength), then ANEC would be violated. Such a result makes one expected to detect ANEC violation for asymmetric perforated systems as figs (**1c**, **1f**, **1g**, **1h**), however in the case of, e.g., maybe hundreds of cavities.

## 3. Post ANEC violation

Indeed, detection of the ANEC violation is the first milestone of generating any shortcut; thereafter we would still have a lot to do. There are papers where after assuming ANEC violation (explicitly or implicitly), discuss about minimizing the energy requirements of a TW (& sometimes wrong claims of removing the CEC violation). They can be classified into three groups: First of them consider CECs violations rounded in space [72,75,76,88], second group ones discuss about delaying those violations & round them in time [9,67,68], eventually the approaches of the third group are too far reaching [17,20,35,36,37], non-standard theories [39], or (mostly) mathematical [15,16].

As an intuitive affair, – something which calculations confirm too – it does not seem reasonable that a mere quantum effect (i.e, ANEC violation in arbitrarily small amounts [73]) could maintain a macroscopic TW for desired applications without any inconsistency [125]. Therefore, we should attack this problem from two fronts: (1) Enlarging the available potentials of producing various types of CECs violating effects. (2) Inventing the models of TWs with the best balances among minimizing energy requirements & other features of an applicative TW. The war in front (1) is "engineering" which its main demonstrating feature is increasing the equipments in the lab (Boxes **2** & **3**); but we have a lot to do in the second front, because "physical" manipulations of the parameters can have the deepest consequences in the model.

## 4. Extraction of NE

Obviously, without a standard theory describing the quantum properties of the fields relating to NE [40] in any manner (e.g., dark matter & energy, within dynamic spacetimes, …) the theoretical frame of the model cannot be considered complete. Although, the most relevant researches to present purposes have been operated on quantum inequalities (QIs) [41,42,43,44,45].

In [44] Ford & Roman discuss about a "bound on the magnitude & duration of NEDs seen by a timelike geodesic observer in 4D Minkowski spacetime (without boundaries) for a minimally coupled, quantized, massless, scalar field in an arbitrary quantum state. That uncertainty principle-type inequality reads:

$$\frac{t_0}{p} \int_{-\infty}^{\infty} \frac{\langle T_{mn} u^m u^n \rangle dt}{t^2 + t_0^2} \geq -\frac{3}{32 p^2 t_0^4} \tag{4}$$

for all $t_0$, where $t$ is the observer's proper time.

Taking into account of state-independent geometrical & state-dependent part of $<T_{mn}>$ to be the source of EM is unnaturally & naively. That's the same for generalizing the Eq. (4) to other massless or massive fields. Other fields have trace anomalies with similar coefficients, i.e., with magnitude of the order of $10^{-4}$. Thus these terms will give a very small contribution to a QI to a of the form of Eq. (4) when $t_0 \ll l$, where $l$ is the characteristic radius of curvature. On the other hand, adding mass will not make it easier to have large NEDs, but one now has to overcome the positive rest mass energy. The effect of including interactions is the most difficult to assess. If an interacting theory were to allow regions of NE much more extensive than allowed in free theories, there would seem to be a danger of an instability where the system spontaneously makes a transition to configuration with large NED".

As an instance of constraints of that theory, consider [31], wherein Morris, Thorne & Yurtsever calculate $r_0 \approx 1$ $A.U.$ for a plate separation of $s \approx 10^{-10} cm$, this wormhole satisfies FR bound. In that model, a typical infalling 3K photon in the CMB radiation, upon arriving at one of the plates, would get blueshifted to a temperature $T \approx 10^{23} K$. In order to traverse it one must go through the plates, that is extremely close to being a black hole, the plates would have to be constructed out of material capable of withstanding Planck energies or more, at last that's unstable. They conclude [44]: If some of the wormhole parameters change over very short length scales, then it would seem from the "tidal force constraints" that tidal accelerations might also change over very short length scales. As a result an observer traveling through the TW could encounter potentially wrenching tidal forces rather abruptly. None of these scenarios seem terribly convenient for TW engineering. Assuming that the stress-energy of the TW spacetime is a renormalized expectation value of the energy-momentum tensor operator in some quantum state, & ignore fluctuations in this expectation value [46], they argue the QIs place severe constraints upon TW geometries.

In the absurdly benign case, for a small "human sized" TW with $r_0 \approx 1m$, FR bound gives $a_0 \lesssim 10^{14} l_p \approx 10^{-21} m \approx 10^{-6} fermi$, or approximately a millionth of the proton radius. Those imply that generically the EM is confined to an extremely thin band, and/or that the TW geometry involves large redshifts (or blueshifts). Also, playing with QIs won't give interesting results. For example human sized values of length lead to very large magnitudes of $|\Phi|$ which positive values give undesirable redshits / blueshifts for a static observer & negative ones make the spacetime close to having a horizon.

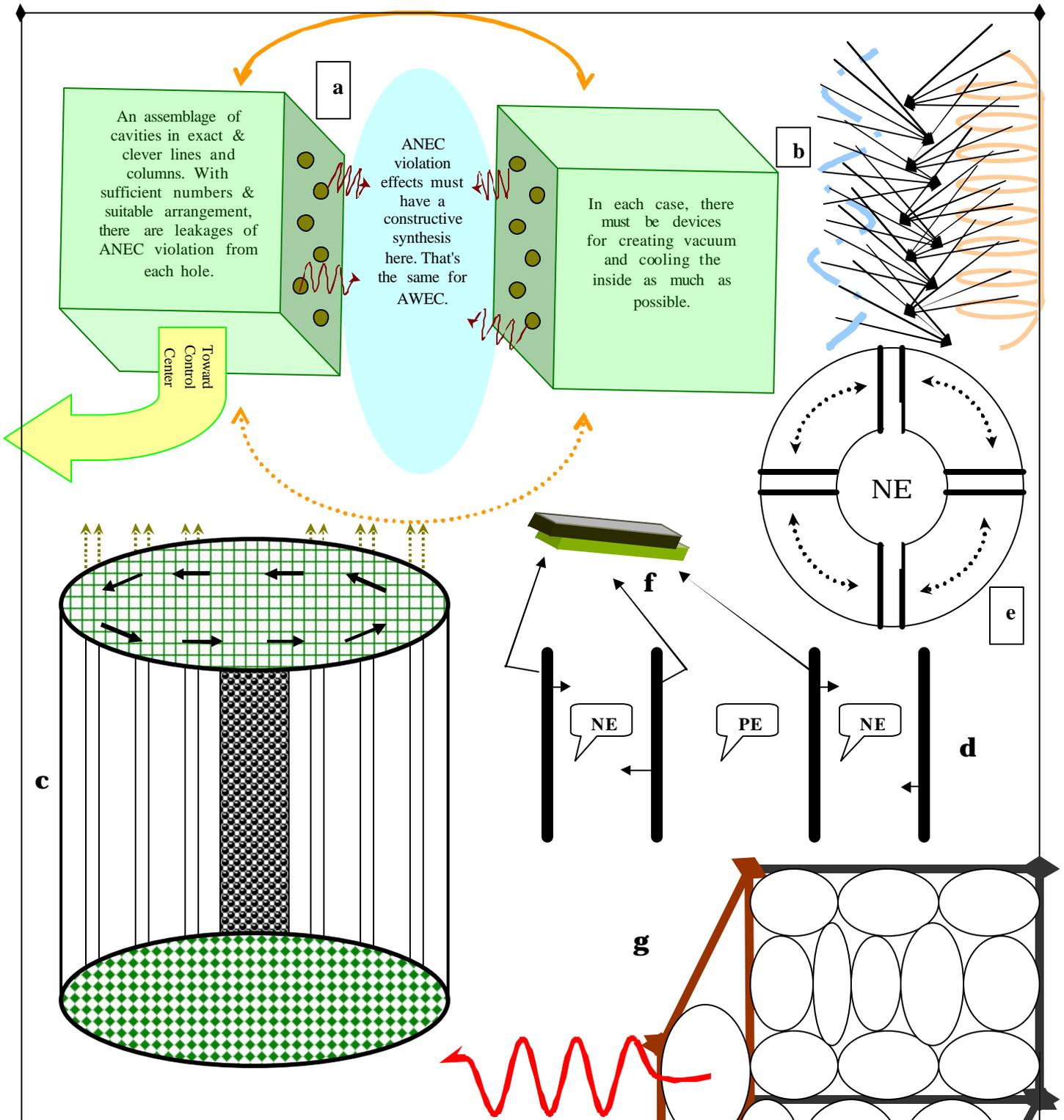

**Box of figs 2**: (**a**) One (non-complete) ring of cases, contained of collection(s) of cavities as in figs (**1c** & **1g**). The fig (**2a**) denotes a ring of cases, but just two against ones are shown and internal cavities are disappeared too. The form of the ring(s) & shape of cases are characterized by technical demands only for more profit. The surrounding inside them have to be as similar as possible to idealized systems (best vacuum, least temperature, flattest surfaces for least dispersive effects, …). The tentacles are connected to a control division to be ordered in specific configurations.

There are arbitrarily numbers of lines and columns, each composed of a collection of perforated asymmetric cavities. The holes depicted can have arbitrary numbers, shapes and locations up to being in favor of mentioned purposes. In the absence of a standard theory of NED, the region of best constructive synthesis (denoted by light turquoise oval region in **2a**) should be specified experimentally (as a good news, the system controlled by fuzzy logic [54], can identify such a region after several experiences [55]). Besides, as a nontrivial result, ordering the pointwise CECs (i.e., DEC, WEC, NEC) violations would be more difficult, because one has to sift the positive contributions of the averaged effects (so, a knowledge of microscopic properties of the averaged integrals seems necessary). One could set up structures of various generalizations of the rings; e.g, different shells, layers, & any ingredient which have such rings. More explicitly, NE of a Casimir system considered as a possible support of spacewarps, happens in a restricted region between the plates. Hence, one needs a more extensive region of distribution of NE for traversability conditions of the metric. In the interacting model of [6,22], we can imagine the interaction at inside region could have induced effects penetrated to outside. Indeed, if prediction of bending the light ray at coming out of an asymmetric perforated system of cavities, would be observed, one can provide a plan for better distribution of NE than before. The plan is as follows: Existence of NED is equalvalent to its gravitational manifestations, i.e., defocusing of photons (ANEC violation) and slower particles (other CECs violations), so as soon as detection of a region having such properties, we must strengthen it by placing the similar regions caused by other collections near together. Various configurations can stretch the NED region, permitting it to be extended without the technical limitations of a simple Casimir setup, hence (longer range of interactions in the fields as a more seminal approach, not being limited to study NED merely at inside regions), the objections of FR [44] to MTY model [31] are modified. Two examples are (gradual) "opening" of plate-like systems (**2a**), and spiral arrangement of systems to pursue an initial detected NED area (point) **2b**). Indeed, one aim is: "not" overwhelming & vanishing of NEDs by (diverging) positive contributions in free space. Also, we should search for stress-energy tensors more general than

$$T_{mn} = (\rho + p)\, u_m u_n + p\, h_{mn} + \Pi_{mn} \qquad (B2)$$

(discussed in [58]) to have a more pleasant and more reachable physics. That is, providing the least restrictions to l.h.s. of (A1) dependent to every appropriate application.

Figs **2c** & **2e**) show two sides of a NE producer tube – definitely, AWEC violating effects – as [22]. The strategies for effective accumulation of NE are used in this case either with more confidence. In (**2e**), the cavity yields NE from two end sides (or one side NE, the other PE). The motion of the plates would be confined to a globally (probably non-complete) circular path (in which the plates in figs (**2c** or **2e**) might be ordered so that after any coming near, find the proper initial configuration *again*) and there are shutters to stop any lossy effects. Also, parallel lines in (**2e**) (might) denote resultant of the assemblies of (**1c**) or (**1g**). If two plates have attractive force on each other representing NE, two adjacent plates might find repulsive effects causing PE somewhere else (**2d**); As a cure for suppress such effects (except of shutters), one may attach an insulator surface to outer side facing to next cavity (**2f**). Fig (**2g**), shows a case contained of tubes as (**2c** & **2e**), they are placed in different orientations in order to give powerful constructive synthesis of interacting fields; i.e., a great flow of NE [124].

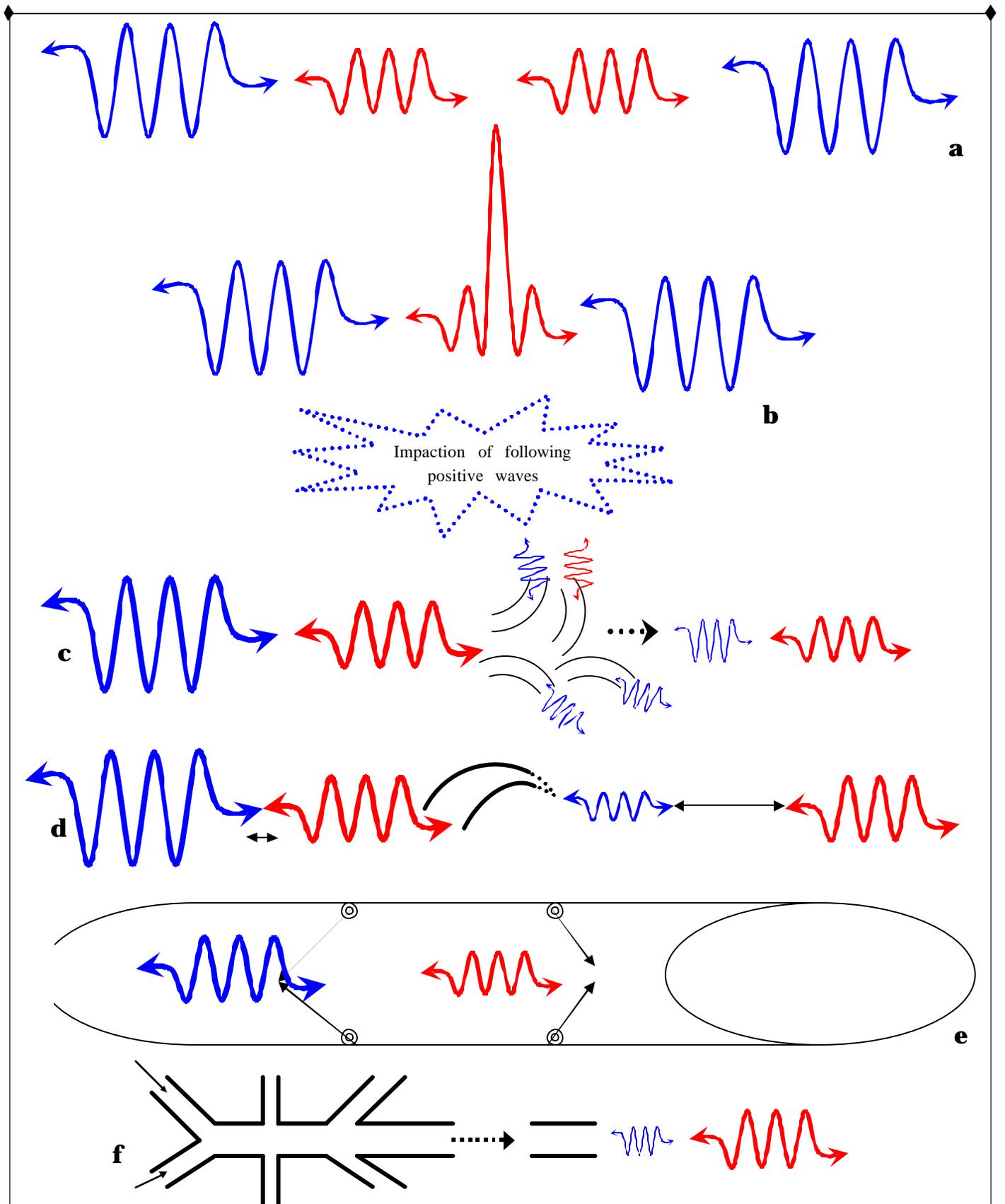

***Box of figs 3***: Engineering scenarios for obtaining more NE. (Blue waves represent NE effects while Red ones represent positive energy (PE) effects);

In Fig (**3a**) two N/P pairs close to each other, negative ones find constructive synthesis and are induced (guided) to another –a deviated– course (e.g., by a gravitationally repulsive mouth of a TW [see § below]), while positive ones are not so (**3b**). This scenario is imaginable for oriented (not-directed) impacts too [60].

In Fig (**3c**), a NE front is adjusted so that lots of tiny TWs are created which disperse following PE. The conclusion is a NE front with a weaker following PE.

In Fig (**3d**), NE makes a dynamic TW so that passes though it and outruns PE (by not allowing positive wave to pass through the TW or have a perfect passage). Of course, the arisen distortions are immediately flattened by PE (up to Planck scales, so a statistical analysis is needed), but any decrease of destructive effects of PE has a great value. In enormous scales, i.e., if one does repeat the tests (**3c, 3d, 3e**) for many times, the probability of more separations between N/P waves, would be increased and even small additional separations can have desired effects.

In Fig (**3e**), the course is decorated by some up and down hooks. The hooks must be lightest stringy structures; e.g., one column of stable metallic hydrogen molecules. The NE effects repel the hooks and continue the path with a weaker strength lost for that work, but during traversing PE, the hooks turn out to be obstructive tools. As a good news, if one designs the end points in an appropriate style (e.g., **S** shaped in which leads to a tight junction), it is expected the PE effects be extremely diminished or even completely blocked.

Fig (**3f**) shows a maze-like collection where at a first glance makes the magnitudes of N/P waves equal through absorbing additional PE by the walls and warming them because of slim course (or increase of velocity?), also by disturbance effects (like sudden twists, etc; mostly for PE), combinations of above tricks in its circuit, and *local* influences like temporary mechanical changing of some parts, gives an output of more NE. Such engineering give contributions to the program of making (B2) more complicated. A strategy which was also followed in Box **2**. (Dependent on extraction of suitable amounts of NE)

§: As [44,61], let $U^m = dx^m/dt = (U^t,0,0,0) = (e^{-\Phi(r)},0,0,0)$ be the four-velocity of an observer who is at rest with respect to the $r,\theta,\varphi$ coordinate system. The observer's four acceleration is

$$a^m = \frac{DU^m}{Dt} = U^m{}_{;n}U^n = (U^m{}_{,n} + \Gamma^m_{bn}U^b)U^n \tag{B3}$$

For the MT metric [62] we have

$$a^t = 0, \quad a^r = \Gamma^r_{tt}\left(\frac{dt}{dt}\right)^2 = \Phi'(1-b/r), \quad \Phi' = d\Phi/dr \tag{B4}$$

From the geodesic equation, a radially moving test particle which is initially at rest has the equation of motion $\frac{d^2r}{dt^2} = -\Gamma^r_{tt}\left(\frac{dt}{dt}\right)^2 = -a^r$. A TW will be called "attractive" if $a^r > 0$ (observers must maintain an outward-directed radial acceleration to keep from being pulled into the TW), and "repulsive" if $a^r < 0$. For $a^r = 0$ the TW is neither attractive nor repulsive.

In that spirit, one can visualize that, up to creation of last entrance mouth to start of being used by passengers, all mouths are repulsive, and in the last stage, they lose repulsion just proportional to the mass of ingoing passengers, –in a certain time – & then the situation converts well.

However, this is not all of the story. There are circumstances in which QIs are violated.

In [47], Hayward states: "Wormhole horizons are two-way traversable, while black-hole & white-hole horizons are only one-way traversable. It follows from the Einstein equations that the NEC is violated everywhere on a generic wormhole horizon. It is suggested that QIs constraining NE break down at such horizons. However, when Pfenning & Ford [48] generalized this method to static spacetimes, the inequalities for static observers were found to depend singularly on the norm of the static Killing vector, which physically encodes the gravitational redshift. For instance, for a Schwarzschild black hole, the inequalities break down at the horizons [48,49]. Thus it may be conjectured that QIs generally break down at trapping horizons. For instance, in spherical symmetry one may take the Kodama vector [50] as the choice of time determining the preferred vacuum state. The Kodama vector has vanishing norm at trapping horizons, becoming zero or null for static wormholes & black holes respectively." (Box **6**)

In addition, there are significant criticisms & counterexamples mainly for applicability of QIs in constructing artificial TWs [51,52,53].

Also, in the model of interacting quantum fields used before [22], it was shown QIs are violated. Further, interaction of matter & spacetime in different polarizations & squeezed states might yield other violations of QIs. (Box **5**)

At last, if one accepts the nature enforces severe constraints on producing macroscopic amounts of NE due to QIs, there is still possibility of adopting a program which physical manipulations in addition of *engineering* tricks can change the situation. See Box **3**.

## 5. Stability

It is possible to arrange the spacetime – by a simple flare-out condition – in which various CECs (mostly NEC) be on the verge of being violated [8,63,64], but since the TW has to be stable, NE is unavoidably necessary for any kind of usage; there are more conditions too [37]. Actually, all features of increasing the NE (except of untractable divergent behaviors; e.g., interaction of traveler's body to huge amounts of EM) are in favor of a TW.

For example, in [65] Poisson & Visser consider linearized radial (spherically symmetric) perturbations around some assumed static thin-shell TWs solution of the Einstein field equations. This permits them to relate stability issues to the (linearized) equation of state of the EM which is located at the TW. One important case in their paper is the energy conservation relation containing the surface energy density $s$, surface pressure $p$, & the radius of the throat (which is a function of time $a \mapsto a(t)$):

$$\frac{d}{dt} s A + p \frac{d}{dt} A = 0 \qquad (5)$$

where $A = 4\pi a^2$ & the parameter $t$ measures proper time along the TW throat. In Eq (5), the first term corresponds to a change in the throat's internal energy, while the second term corresponds to the work done by the throat's internal forces. The TW is stable if & only if (at least until the nonlinear regime is reached):

$$\frac{2M}{a_0} + \frac{M^2/a_0^2}{1-2M/a_0} + (1+2b_0^2)\left(1 - \frac{3M}{a_0}\right) < 0 \qquad (6)$$

On the other hand, if one treats $b_0$ as an externally specified quantity, stability may be rephrased as a restriction on the allowable radius ($a_0$) of the assumed static solution.

The analysis indeed shows that once the effect of the TW's gravitational field is included, stability implies that large TW ($a_0 > 3M$) are stable only for $b_0^2 < 0$! [there are several known examples of such exotic $b^2 < 0$ behavior: in the test field limit, the Casimir vacuum [30] between parallel plates is known to be of the form $T^{mn} \propto diag(-1,1,1,-3)$. Integrating over the region between the plates, the 3D surface stress-energy takes the form $T^{ij} \propto diag(-1,1,1,)$. In this case $b^2 = \partial s/\partial p = -1$. A similar argument shows that $b^2 = -1$ for false vacuum, for which $T^{mn} = \Lambda g^{mn}$, where $\Lambda$ is a constant].

Briefly, having sufficient NE with smooth distribution & suitable qualification of behaving in spacetime, leads to desired stability.

Since possibility of being high-sensitive of parameters corresponds to severe perturbations in physical manifestations & disappointing probabilities (e.g., a dangerous amount of PE can annihilate a TW [47]); combining two approaches of [72, 82], the concept of "practical stability" along with secure margins would be introduced in the following.

## 6. Features of a desired shortcut

### a. TW element

Let us deal with the suitable corrections to a TW metric [62] to become practical.

• The metric certainly has to be dynamic [66]. The lifetime of a TW must be middle of two extremes. (*1*) The TW shouldn't be long-lived. The long-lived dynamic TWs can be almost considered static, equivalent to unpleasant implications for the geometry and energy demands. (*2*) The TW shouldn't have small lifetime neither, mostly because of tremendous tidal forces. For each operation, the lifetime of the TW, is dependent to special parameters of that travel; i.e, distant, goal (exploratory, civil, ...), the status of passengers (humans, photonic signals [69], ...), etc. On the other hand, due to the QIs, NEDs are time-dependent, so that's the same for any related geometrical structure. Also, particle creation in such non-static spacetimes should be noted [70].

• The metric must be generally asymmetric [74]. The assumption of spherical symmetry has been using mostly because that extremely simplifies the calculations by making a more consistent study in Schwarzschild coordinates; but if your intension of researching on TWs is more than theoretical studies or possible applications with huge orders (e.g., proper distance length to the throat [80]), your main choice would be nothing but an asymmetric geometry with severe consequent mathematical difficulties of GR. (Loosely speaking, the motto is: harder math, easier physics). Such modifications must involve to change the operation range of TW from connecting two arbitrary regions of spacetime, to two limited points with better conditions for distance of an observer to the throat, through numerous cut-offs (Box **4**). Regarding TWs in the most general geometries and topologies, gives one more possibilities to have a reasonably consistent adjustment of involved parameters, i.e., giving up spherical, even any kind of symmetry & $R \times S^2$ topology. It must be noted, nothing – definitely, rate of evolution & Lorentzian topology change – should lead to causality disconnecting (of valuable information) in the process.

• The metric should use back-reaction effects. The only known tractable one is the addition of the charge to a MT metric [81]. Therefore, if there would be an arrangement of imposition of the charge (certainly temporary) inside the NED maker equipments without destructive interference to other surrounding (Box **8**), one is expected, in special locations and time, charge effect would be able to dramatically change the geometry of a TW, leading to its enlargement, (In fact, very more than the symmetric case of [81]). In present strategy, (at least) charge back-reaction can decrease the unpleasant sensitivity and its consequences of a model, along with thoughtful functions, smeared by dynamics and asymmetry, Eq (7).

As a matter of fact, generalizations of MT metric up to now, have been only mathematically well behaving modifications away from MT; but it seems a practical metric must be very more complicated. Perhaps such an ideal metric wouldn't have a specific analytic form. As a proposal, in the most basic level, we should start from $ds^2 = g_{mn}dx^m dx^n$, then considering the most fundamental element of a TW – throat –, and specify all other desired parameters as sets of numerical approximations in which, the proper distance length should have a better situation by limited intervals of the integration for $l$ rather than in the Ref. [82]. Also, vectors other than the 4-velocity of a (null) particle that is directed *radially* inwards/outwards, i.e., $k^a = (\sqrt{-g^{tt}}, \pm\sqrt{g^{11}}, 0, 0)$ must be taken into account as the proper course of the travelers. Such scenario can be cast into a global coordinate design for any operation. Similar to directing e.g., a Cruise rocket, one can write a program as a spacetime extrapolation for the travelers and attached shortcut maker equipments (representing intelligent behaviors by specific lifetime and general options only dependent to expectation of the user). The main difference is, in former case, starting by Newton's equations for a trajectile, some restrictions are imposed to the system to use the best path, e.g., by gained information of thermo sensors; but in the latter, starting by semi-classical GR equations, employing 'variational principles' (to not waste additional energy & distance), different types of constraints are imposed to the system. Those are gained by adjusting the properties of an appropriate TW (redshift & shape functions – in the form of $b_n$ & $\Phi_n$ as path-finder series – , asymmetry, charge, dynamics) and are dependent to what the inside is (human, goods,…), locations of would-be-connected ports, also taking into account the perturbations caused by interferences & security problems. Such a strategy can be governed by a nonlinear control theory used in the field of artificial intelligence (fuzzy logic [54] seems greatest). Therefore, the TW searches allowed paths (& digs the spacetime) to find the best one for its destination, corrects its mistakes, and similar to an intelligent entity (or an internet search engine), its demonstrations are unpredictable – because its choices are quite general – but within regarding the best efficiency to the user.

Indeed, after creating a region of spacetime which is not simply-connected by ANEC violation, one can assume a TW metric that uses all the known options for its survival and efficiency.

Other possibilities should be considered too. For example, as pointed out in [83], if spacetime has to maintain a well defined spinorial structure on it, then the creation of wormholes should occur in 'pairs' [84]. Therefore, after a pulse of ANEC violation, much attention is needed to probable disturbances, like such phenomenon.

Also, nature would render mechanisms for causality survival fortunately, as an instance, in [85] by schematic representation of a single (chronology respecting) TW and identifying two timelike lines that are separated by a spatial jump $L$ and timeshift $T$, with $T \ll L$, Visser shows if you combine a system of $N$ identical TWs in otherwise flat Minkowski spacetime – for simplicity assume that all TW mouths are at rest with respect to each other – gravitational back reaction makes the semi-classical quantum gravity approach unreliable; see the fig 1 therein.

**b. WD element**

My personal interest is studying the spacetime ways of transporting as much as possible close to a mere TW; but as you'll see, the idea of combination of a TW to the other candidate of FTL travels in the literature, i.e., WD metric [78] can cause better results. The main motivation of such inclusion is the energy problem discussion; then the possible lessening or even removing of time dilation by the bubble model – note there is no limitation to velocity of a warp bubble (however for technical reasons, preventing horizon formation or high needed EM, it is assumed $v < c$) and such system can make the time dilation of the trip infinitesimal for the observers, along with the TW background geometry contribution in addition of other propulsions like carrying EM by inside spacecraft & of course the WD metric itself by finding appropriate squeezing effects, – and presence of natural tidal forces in a WD metric (at least in the immediate vicinity of the spacecraft, provided a big enough bubble) are further reasons of that decision.

The outcome may be (loosely) considered as a subclass of TW spacetimes; However as shown in Box **7**, one can visualize it as a "virtual pregnant worm" (VPW). There are reasons in which make one not to have a strange feeling to that terminology. First, the base geometry is of a TW, *then* a WD is added by a surgery. As shown by Van Den Broeck [86], one can make a roomy bubble by some changes in the parameters. This stage only needs minor EM, and the necessity to EM can be much reduced [88] further. Hence, one has a microscopic TW+WD (*note*: not in Planck scales) and can identify the destination with a high freedom. In addition, the role of artificial intelligence in the form of fuzzy control of the tools for engineering demands, makes the geometry to have unique demonstrations for every operation; similar to an alive worm!

Therefore, the program is as follows: The systems gets ready to concentrate a region of EM in a least possible magnitude, controllable for the tools. Less thin that region, more capable of Van Den Broeck trick and less necessary EM for maintaining following bubbles but more density of EMs (unpleasant NEDs will be relaxed later, see next parts). However, it seems that is most an engineering challenge (& more amounts of EM, less need to such tricks); e.g, that region must be comparable to proton radius to be handled by femto-lasers or be flowed through a nanotube or any other appropriate (cutting edge science) mechanism. Box **8**.

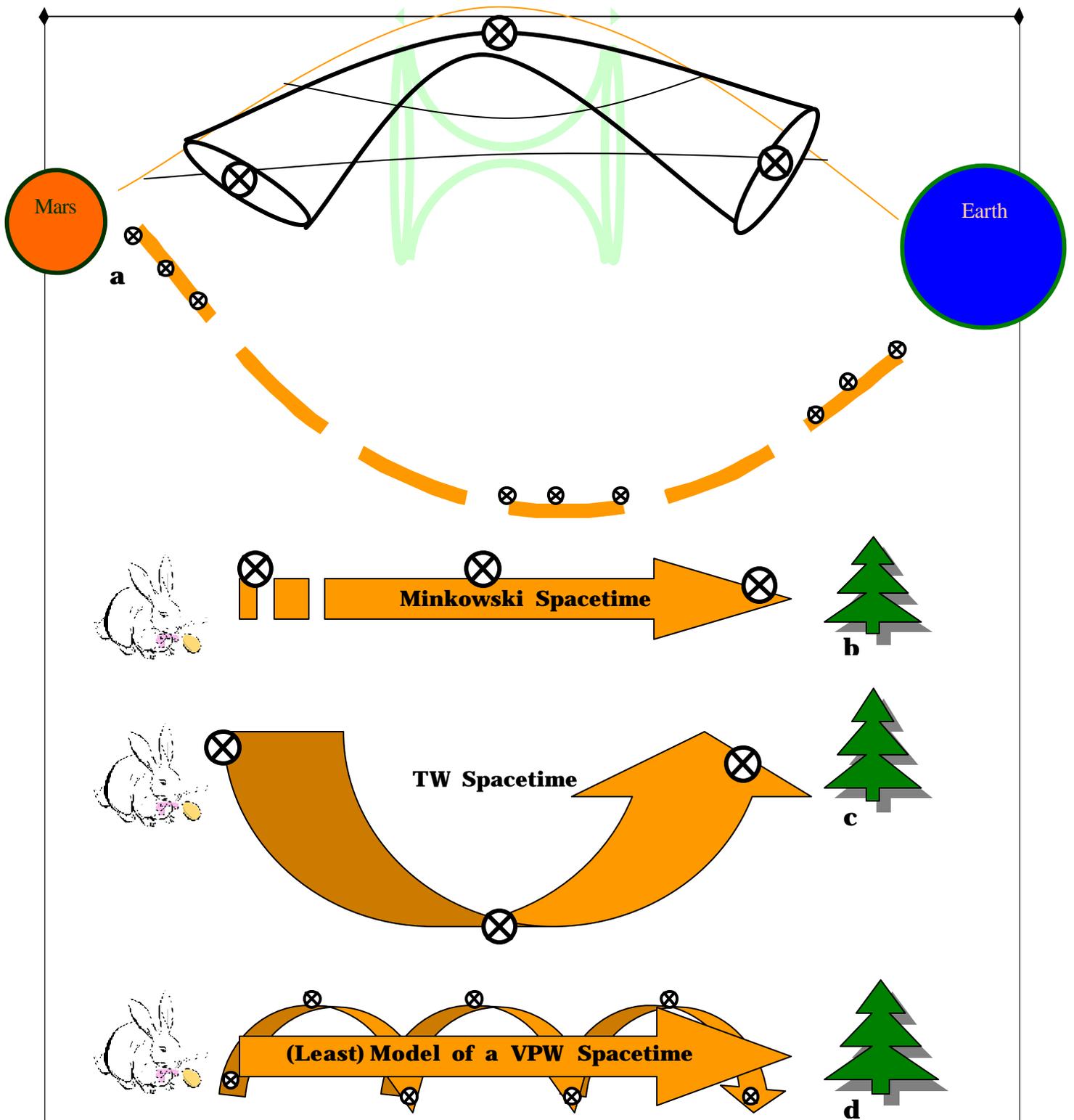

***Box of figs 4***: (**a**) It seems reasonable, "finite" amounts of NE, could cause a TW connecting a "finite" set of points of spacetime (both *mathematically* and physically). In other words, in studying intra-universe TWs, the 'universe' under consideration should be quite in the usual scales & not of cosmic dimensions (or connecting two points *arbitrarily*). Therefore, the concept of limitedness of effective range of operation of a TW modifies the asymptotic flatness (via reconsideration of the definitions of the flaring-out conditions [see (A3) & (A4)]) for the practical applications, dynamics of metric, ANEC violation with finite interval of integration, i.e., different from $(-\infty, \infty)$, etc. So the strategy is: Go to large distances by 'partial' effecting on spacetime *not* "an overall distortion".

Other three figs show approaches for faster achievement to a destination. First rabbit increases her velocity in a Minkowski spacetime to reach to tree (**b**). Second rabbit passes through a general (somehow static) TW system. Indeed, a kind of shortcut is used in arriving to tree (**c**). Nevertheless, the way third rabbit chooses cannot be called a shortcut model (in traditional manner). Instead, she does some *leaps* to avoid of both high curvatures (and consequently severe energy requirements) and slow flare-out (as is [73,89,90, and almost 82]).

By means of a balance among dynamics of every jump, its geometry and topology, suitable choices of shape and redshift functions, and auxiliary back-reactions, one may expect third rabbit leaves the first one behind in a racing by temporary modifications in spacetime, equivalent to sudden peaks in her velocity; similar to a doper racer.

As a matter of fact, more studies need to be done on probable time confusion in (**d**), but "as a conjecture", a combination of Novikov consistency conjecture plus Hawking chronology protection conjecture [3] might act against conversion of a VPW to a closed timelike curve.

Therefore it seems reasonable by intelligent chains of (TW+WD) systems, presenting essential energy requirements and spacetime configurations, a transition by spacetime shortcut can occur at least for non-dreamy distances; sending humans or goods to earth-like planets in the solar system, like Mars, Europa, etc.

*Remark* : To circumvent "the fact that at 300 times FTL impacts with asteroids comets supernovas or Black Holes also photons of background radiation Doppler blueshifted to energies of the radiation synchrotron impacting the Warp Bubble would disrupt the Warp Field as for TWs", one can focus to identifying some front points (denoted by tiny crossed circles) along with adjusting secure margins to avoid damaging the VPW. That must be done by the common detection equipments of a spacecraft using the VPW for its propulsion; similar to an intelligent guided missile which avoids of hitting to mountains, anti-missiles, etc. Besides, because of different situation of the throat(s) in *dynamic* and static TWs [63], "in order to use a TW from a realistic point of view to travel from Earth to Andromeda Galaxy for example you must have one of the mouths here and the other mouth in Andromeda Galaxy of course ... and who creates the other mouth in the first place?", locating the other mouth *during the operation* and convincing to lower scales such as Earth-Mars (instead of Earth - Andromeda !) can be an answer to that problem, at least for the first generation of VPWs. At last, the drawback of "having an event horizon (Penrose diagrams, Kruskal-Szekers or Boyer-Lundquist coordinates) and lack of directing the creation of a mouth in a desired place distant from Earth", seems to be removed by choosing an element of VPW as the *subluminal* WDs & exact locatings like military missiles, recent discoveries on Mars, etc [126].

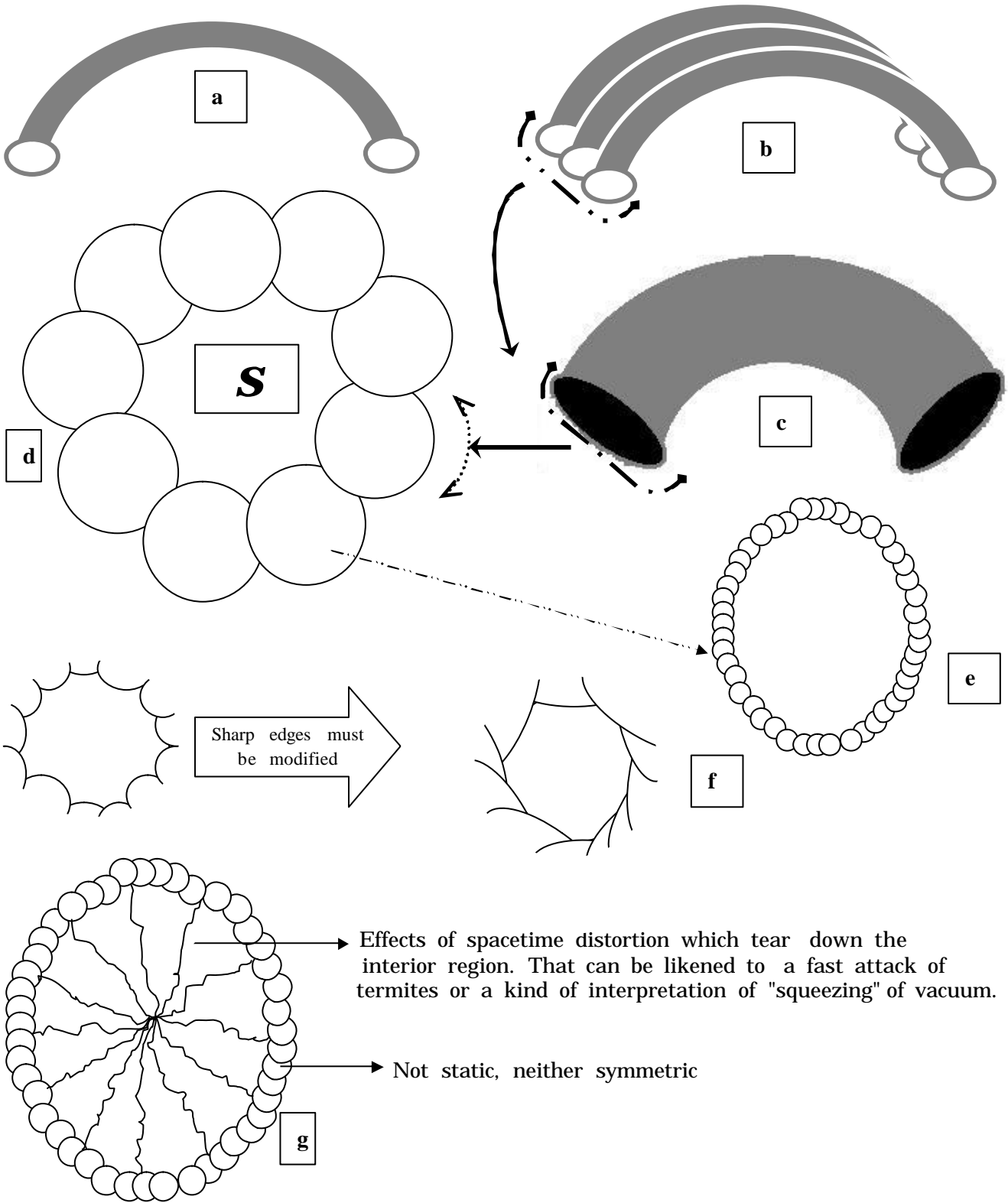

***Box of figs 5***: First, there is a TW (**a**); If we arrange e.g., three of them near to each other (**b**), it seems to be able to gain one with bigger size in mouth and throat (**c**).

Now, the question is: What would happen if we extend that settlement to a complete ring (or any closed curve) of mouths which none of them are completely distinct (i.e., have common regions)? (**d**)

Definitely, what would be the destiny of the middle region (**S**)? Is it possible in principle to create a large TW, by the mentioned arrangement of numerous tiny ones (**e**)?
Can it improve the energy requirements? and what difference is between the static and dynamic TWs in such configuration?

Without an explicit interpretation of (island–like) PED regions, (however those probably cause instability effects to a required shortcut), one can think to cleaning them by a 'vacuum cleaner'–like device (temporarily). The NE-issuing tube (Box **2**) *cleans* those *dirty* regions of PED.

If resonance of the effects of smaller TWs and converting the system to a bigger resultant, be possible, that can yield a great result; i.e., the creation of regions causing squeezing of vacuum and NE.

*Notes* : Breakthrough in the behavior of tiny TWs with oriented squeezing and polarization of vacuum as (**5g**), would give major contributions to any influence on spacetime which cause macroscopic demonstration (and as a good news, there would be less necessity to artificial pumping of EM). That would be done through a special imposition of thoughtful regions of NED (created by the very curvature of spacetime see [16,91,97]) & constructing (not necessarily self-maintained) TWs. As a suggestion, during a surgery the system applying the fuzzy control, would find the best order after some repeats.

However embedding diagrams for dynamic TWs are misleading, but above figures are for better intuition. Also electromagnetic scanning [56] may clarify the shape of process. That's similar for the Boxes **7**, **8**.

**Box 6**: Some remarks on a desired geometry:

If one can construct a TW, nothing would be more vital than the comfort of the passengers. Therefore, there always must be a confident secure margin. That finds more importance when one does a "balance game" in a time-dependent geometry by the boundary of, having or not having the essential conditions which define a TW. (Maybe, on (in)equalities of geometrical properties, one can perform a "balance game" and uses the advantages placed on the borders; that's the case for previous discussions too.)

Indeed, the concept of the horizon has a great potential to the energy improvement. This is based on two facts; breaking QIs on the horizons, and providing suitable conditions for "volume integral" theorem of total amount of EM proposed by Visser, Kar & Dadhich [72,92]. One may consider it as a generalization of cut-off imposed versions of TWs energy discussions, started by 'absurdly benign' solution of MT [62].

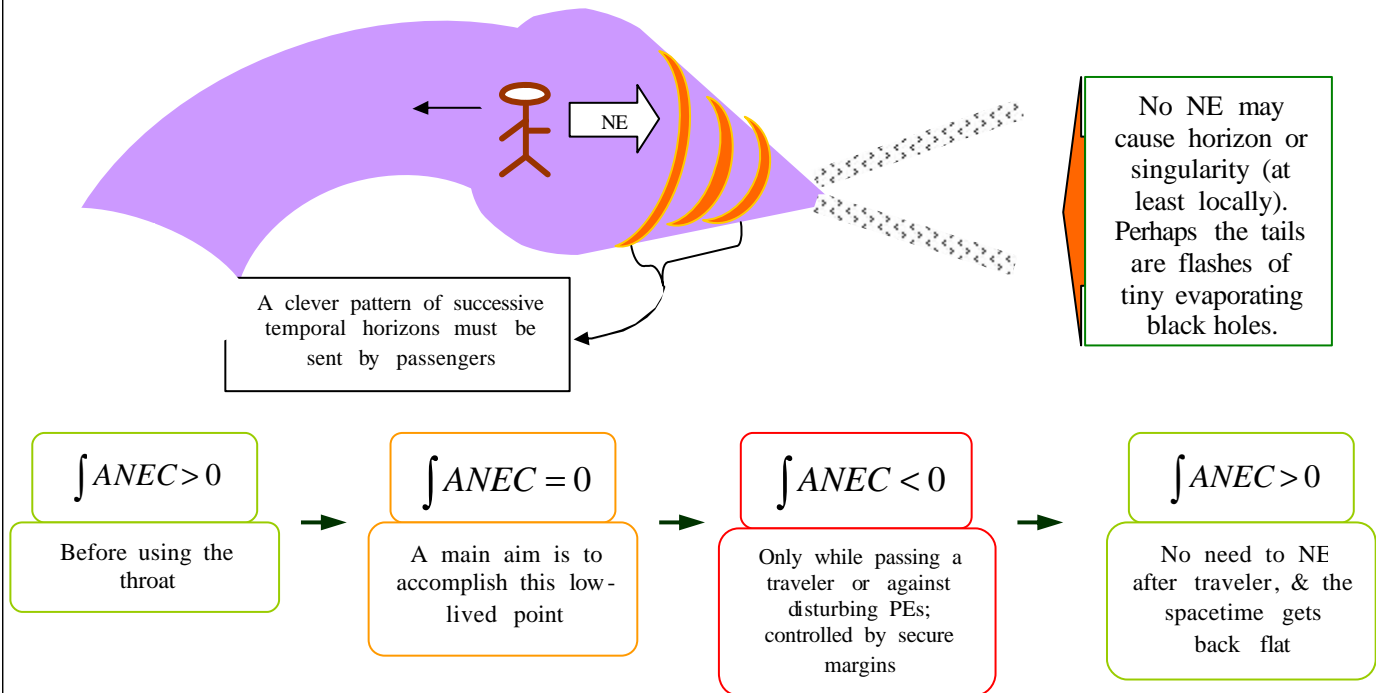

A clever pattern of successive temporal horizons must be sent by passengers

No NE may cause horizon or singularity (at least locally). Perhaps the tails are flashes of tiny evaporating black holes.

| $\int ANEC > 0$ | $\int ANEC = 0$ | $\int ANEC < 0$ | $\int ANEC > 0$ |
| --- | --- | --- | --- |
| Before using the throat | A main aim is to accomplish this low-lived point | Only while passing a traveler or against disturbing PEs; controlled by secure margins | No need to NE after traveler, & the spacetime gets back flat |

If one assumes the two-way horizons of TWs, by breaking QIs on the horizons either [98], dynamic TWs would then encounter to an alternating cleaning the ergoregions, (for more insight to interconvertiblity or bifurcation of such (double) trapping horizons, see figures, e.g., at [93]), one might deduce the concept of event horizon is totally dependent to presence of EM. Therefore, that seems an appropriate idea, which by creation / annihilation of numerous horizons in the lifetime of a dynamic TW [99], one could gain a dominant control on less constraints of NE [100]. That is more interested to subtle experiments when you meet 'small' needed amounts of NE. Note in huge amounts, there are two destinies for a dynamic TW; converting to a black hole or an inflating universe! [47,93,94,95].

However, due to [64], temporary suspension of the violation of the NEC at a time-dependent throat also leads to a simultaneous obliteration of the flare-out property of the throat itself [63]. So, we must pick up a formalism containing limiting behavior to that problem; thus, note Eq (B5) represents the arbitrarily small total amount of EM, provided $e^{g(r_0)} \to 0$, where is the throat of the TW. The limit actually refers to a sequence of TWs.

But in an asymmetric version, up to a flare-out corresponding to be on the verge of violating the NEC – ideal condition equals to zero energy density, more values require faster return to normal situation – any conduction of perturbing effects e.g., diverging fluctuations at particle horizons for the case of WD with $v_s > 1$ [101], away from passengers to elsewhere of spacetime would be useful.

Now the second fact: The skeleton of the metric is of Kuhfittig [82], subjected by an imposition of VKD [72]. It means when one considers

$$\int [\rho + p_r] \, dV = 2\int_{r_0}^{\infty} [\rho + p_r] 4\pi \, r^2 dr \qquad (B5)$$

as a suitable measure of the 'amount of EM' required to maintain a TW, the spatial slices must be the same as of Eq (7). Therefore the 'proper volume' of the region of EM in the Kuhfittig III model [73] accepts further reduction (see p. 23 of [80]), and "an extensive region of NEC-violation" up to a cut-off to both satisfy QIs & *radial* tidal constraints necessary for human traversal [102] meets a new dynamical diminution.

As a matter of fact, there seems a deep inconsistency between a real TW & severe restrictions of QIs (in all models, e.g., MTY [31], Kuhfittig III [73,125], cut-off imposed versions, etc). Regarding continuity implications (to functions & derivatives), the above scheme tries to circumvent the problems such as, unsafely being close to a Schwarzschild wormhole, causality violating discrepancy between the traveler's clock & of an other observer in the external universe (i.e., an appropriate traversal time), and enormous blue/red-shifts. Therefore, any theoretical framework more relaxed than QIs would give new promising consequences.

# 7. Prototype of an Experimental Model

The metric describing a VPW, is a suitable combination of the terms in previous attempts to improve the stress-energy implications of TWs and WDs, in addition of fuzzy control [54] of the system. As soon as the birth, a VPW works on five principles:

1) There is a destination, 2) It has to take something through itself to there, 3) Time of operation must be less than a similar transmission through a Minkowski spacetime, 4) So, it has to dig and warp the spacetime by the most efficient and secure methods to both passengers and creating tools, and note because of non-linear structures, one must be exact in choosing parameters. The effects like longing the course or its non-sufficient reduction, turning around oneself and so on, are probable and decrease the efficiency. 5) At least during initial moments, the entrance mouth has to be macroscopic, as a curved spacetime in respect to any proper user.

Let us visualize a program for creating a VPW; the metric of spacetime can be written as:

$$ds^2 = \Omega(t)\left( -\exp(2l(r,t,\boldsymbol{q},Q))dt^2 + \Xi\ [\exp(2m(r,t,\boldsymbol{q},Q))dr^2 + K^2 r^2 (d\boldsymbol{q}^2 + \sin^2 \boldsymbol{q}(d\boldsymbol{f} - \boldsymbol{w}\ dt)^2)] \right) \quad (7)$$

Now, several introductions:

**I:** *Time factors*: The $\Omega(t)$ is a conformal dynamic factor discussed in [67,68], $\Xi$ is another acceptable factor which is the same analyzed in [61] if $\Xi = \exp(2ct)$ or in [38] if $\Xi = R^2(t)$, it should be noted to the form of $\Xi$ and more effectively, the terms containing of $\Xi \times \Omega(t)$ to avoid of unwanted properties like unpleasant enlarging the radial proper distance between the TW mouths (in addition of the size of the throat in an inflating model [61]), as well as, in certain intervals, there are $r$ dependences of $\Xi$ corresponding to $B$ factor in Eq. (8), used in a forthcoming surgery. Also, all the relations must be continuous in respect to $t$ variable, up to not losing any valuable information.

**II:** *Essence of a surgery*: The outlines can be considered as:
The surgery is the combination of a TW to a Van Den Broeck–Alcubierre WD metric of the form:

$$ds^2 = -dt^2 + B^2(r_s)\left( [dx - v_s(t)f(r_s)dt]^2 + dy^2 + dz^2 \right) \quad (8)$$

leading to a VPW. Generally, it can be inspired from where Alcubierre started [78, pp. 2–4]. One has:

$$ds^2 = -d\boldsymbol{t}^2 = g_{ab}dx^a dx^b = -(\boldsymbol{a}^2 - \boldsymbol{b}_i \boldsymbol{b}^i)dt^2 + 2\boldsymbol{b}_i dx^i dt + \boldsymbol{g}_{ij}dx^i dx^j \quad (9)$$

along with parameters different from of a warped bubble; i.e., the lapse $a$ function that gives the interval of proper time between nearby hypersurfaces as measured by the "Eulerian" observers (those whose four-velocity is normal to the hypersurfaces), is not a constant as assumed before ($a=1$), and the shift vector $b^i$ that relates the spatial coordinate systems on different hypersurfaces cannot be confined to one axis [and as a symmetric case one can have $b^x = b^y = b^z = -v_s(t)f(r_s(t))$], particularly the 3-metric $g_{ij}$ of the hypersurfaces does not equal to $d_{ij}$ anymore; but that represents the spatial part of Eq (7), of course without $r$ dependences of $\Xi$. However, it remains an open question, since the 3-geometry of the hypersurfaces is TW element with high discrepancy of the involved scales – a roomy bubble along the throat of a TW, diminished by engineering requirements – , can we deduce the information about the curvature of spacetime will not be contained in the extrinsic curvature tensor $K_{ij}$? Or if that cannot be *completely* subjected by TW background, how would the 3D hypersurfaces embed in 4D spacetime? Also, more complexities infest on the junction point. Due to the fact that warp bubble has a hole, the fig 1 in [86] need to be extended to more regions to describe a VPW; e.g., region V for TW element and more manipulations against unpleasant *densities*, etc.

Any version of a WD must be "subluminal" to give a suitable geometry. Because, bubbles with $v<c$ do not cause any horizon (and actually, diverging vacuum fluctuations) [103], require less EM [79,86], are casually connected, and have a slow motion of EM. Also, modifying WD geometry by corrections as of Krasnikov tube [104] would help to casual problems. In addition, according to [103], there is more chance of to compatibly circumvent the QI and quantum interest conjecture by a trick. Besides, $B^2(r_s)$, $f(r_s)$ & $v_s(t)$, must be quite general, i.e., not restricted just on the $x$ axis, have more dependences as of $l$ or $m$, along with further correspondence & manipulations (e.g., terms mixed to $\Xi$), and more complicated definitions of $B^2(r_s)$ [88]. Other point is the main attention to the high energy *density* in the metric. In fact NEDs must find sufficient relaxed distribution in WDs as of TWs. To that end, one has to apply suitable dynamics, asymmetry, the local backup like Coulomb forces and combinations of all them as was so in TW element. For instance, in a QIs breaking model of Krasnikov [88], giving up the spherical symmetry reduced the required energy by 35 orders (although still huge).

**III:** *Backreaction of charge*: in [81] authors consider the metric of

$$ds^2 = -\left(1+\frac{Q^2}{r^2}\right)dt^2 + \left(1-\frac{b(r)}{r}+\frac{Q^2}{r^2}\right)^{-1}dr^2 + r^2\left(d\mathbf{q}^2 + \sin^2\mathbf{q}\ d\mathbf{f}^2\right) \quad (10)$$

as a model for a "charged wormhole". Since they consider the static case, there is no radiation by the fields; and due to being positive of the components, no horizon formation occurs. That means the addition of charge might change the TW but will not change the spacetime seriously. Also, they conclude a charge of $Q \approx 3 \times 10^{16}\ C$ would forbid formation of a TW with the throat of 1m radius. Now some points should be noted: By letting coefficients to have arbitrary signs as in the case of Eq (7), one can principally derive more effective influences on spacetime; but in contrast to MTY model [31], – where outside of the plates is a classical radial Coulomb field with $rc^2 = t = p = Q^2/8pr^4$, which produces a Reissner-Nordstrom geometry –, dynamics & asymmetry of the proposed model (see fig **8a**), prevent of severe stresses to the plates in addition of the difficulties of force balance at the plates or forbidden separation either. On the other hand, according to the terms containing $\dot{Q} = i$, we see more ability of engineering in the electric tools as illustrated in Box **8**. It originates of our technological domination on electromagnetic effects. In that spirit, – regarding enormous involved values of charge – a process in a *reversed* direction might promisingly give significant contributions. For example, as an analysis in [80], Kuhfittig model I [89] needs tremendous precision for fine-tuning; but having armed that highly sensitive model to a charge backreaction, would reduce the required precision through every one magnitude of order increasing a Coulomb by $60/16 \approx 3.7$ magnitude of order for a 1m throat "which satisfies QIs" & $30/16 \approx 1.8$ magnitude of order for a proton-sized throat. Therefore, there would be plenty of good news (at least on some spacetime regions) in dynamic asymmetric & probably QIs breaking geometries, attached by the charge effects. Also, such powerful backreaction effects might provide situations –by complicating a simple gravitational framework– with high probability of trapping the non-geodesics rays. See also [124].

**IV:** *Attempt to go beyond a spherical symmetric line element*: Although the ideal aim is a complete asymmetrization (see Appendix), but for this stage, we can content ourselves – to start experiments – with a less simplifying framework; i.e, "axially symmetric" TWs. The last part of Eq (7) belongs to the rotating *axially* symmetric case discussed in [9,105]. That generalization makes the WEC violation considerably less severe than MT metric. However, a rapid rotation will result in a reduction in the WEC violation in the *spherically* symmetric case, it also adds to the constraints of lateral forces. On the other hand, there are more types of generalizations. For example, totally antisymmetric torsion actually promotes the CEC violation at the throat, but helps to lessen it away from the throat by generating twist. Besides, other changes are possible [106] as may be observed in the WD metric [79,86,87,88].

**V:** *Other dependences of $l$ & $m$*: A main concentration is on the reasonable parametrization of shape and redshift functions. Those are dependent to four parameters (to relax the NE *densities* by rounding them in space & time in a complex streaming). The involvement of $r$ is trivial, the cases of $t$ and $q$ were reported in [90,105]. As a guidance, as expressed in [82] & confirmed in [102], one has better to start the study with 'logarithmic' functions for $l$ and $m$ (mainly to normalize infinite magnitudes). Also, all the relations must be continuous in respect to spatial parameters up to being safe to the travelers.

**VI:** *Additional considerations*: As you see, we have a metric to start, attached by many constraints. The arisen constraints come from initial assumptions, and more complications, more such constraints; Some of them are:

• In order to avoid a singularity on the axis of rotation, and the $q$-derivative of the functions $\sqrt{\Omega}e^l$, $m$ and $K$ must vanish for $q = 0, p$.

• If we define $b = r(1 - e^{-2m})$, then $b$ must satisfy the inequality $b \leq r$, $\forall r$, with the equality valid only at the throat.

• $\left.\frac{\partial b}{\partial r}\right|_{r_{th}} < 1$ (flaring out condition).

• $\left.\frac{\partial b}{\partial q}\right|_{r_{th}} = 0$ (to ensure that the curvature scalar is nonsingular at the throat).

• $\sqrt{\Omega}e^l$, $\forall r$ must be finite and nonzero (to ensure that no singularities or event horizons are present).

• The angular velocity must be independent of $q$. (For a macroscopic TW [107]).

• In the case of "axially symmetric and stationary" spacetimes [105], considering Eq (B2), there are restrictions on the r.h.s. of (A1) [58]:

$$G_{12} = 0 \quad (11a) \qquad G_{11} = G_{22} \quad (11b) \qquad G_{03}^2 = (G_{00} + G_{22})(G_{11} - G_{33}) \quad (11c)$$

– a perfect nonconvective fluid as a source –

$$G_{00} + G_{33} \geq 2G_{03} \quad (12a) \qquad (u_0/u_3)^2 > 1 \quad (12b)$$

– a fluid with anisotropic stresses as a source –

Kuhfittig has shown [105] with a special choice of $m$ and $l$, one can sufficiently satisfy the conditions (11a) and (11b). Therefore, it seems reasonable by other choices for $m$ and $l$, one can satisfy more conditions. However, Eq. (7) is either dynamic & may circumvent above restrictions by a perfect fluid admitting a proper conformal motion as a source [59], the role of time factors is significant herein, mostly for TW element of VPW.

Above statements restrict the proper path of the travelers but do not seem to forbid all operations; whereas the above restrictions can be subjected to the geometry only at initial moments and (or) by suitable (radial) cut-offs of the stress-energy tensor, one can construct a solution using required junction conditions to external solutions.

Therefore, since more complexity, more properties of the arbitrary courses, but less numbers of those courses, the path of a passenger (like "dancing") would be confined on a 3D or 4D manifold (regarding other factors too) with the least probability of encountering the event horizon, unbearable tidal forces, and other unpleasant features.

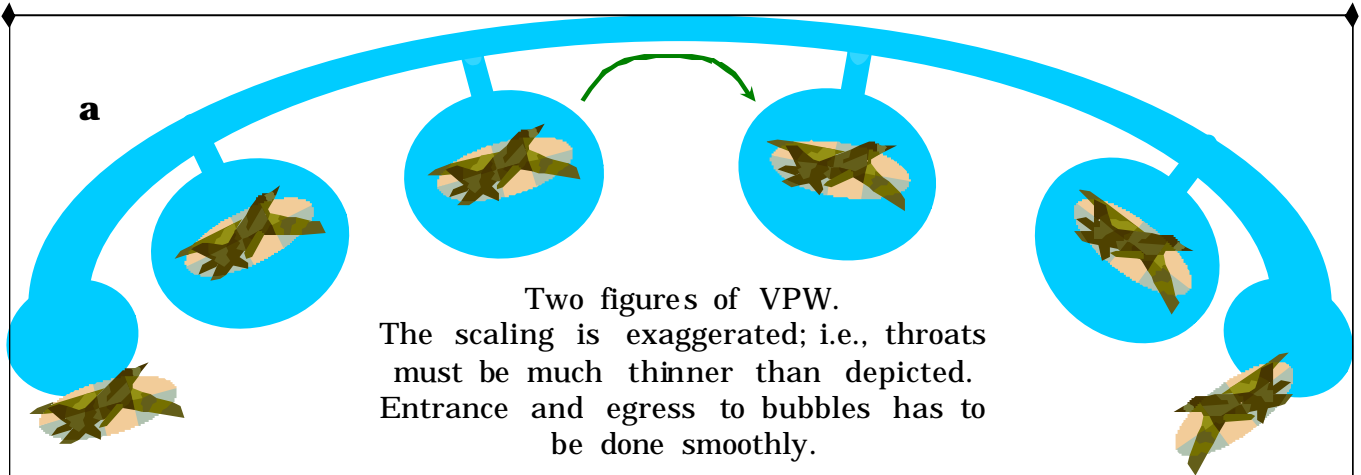

**a**

Two figures of VPW.
The scaling is exaggerated; i.e., throats must be much thinner than depicted. Entrance and egress to bubbles has to be done smoothly.

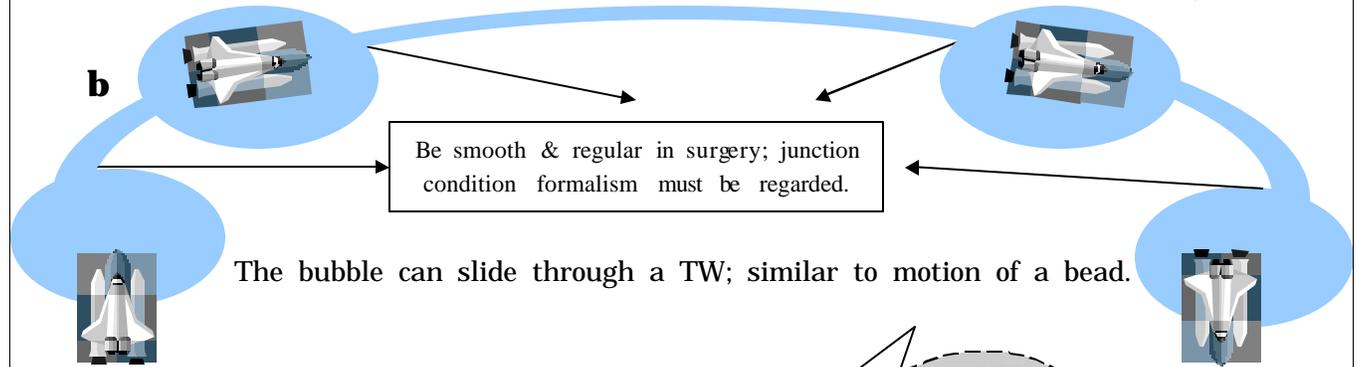

**b**

Be smooth & regular in surgery; junction condition formalism must be regarded.

The bubble can slide through a TW; similar to motion of a bead.

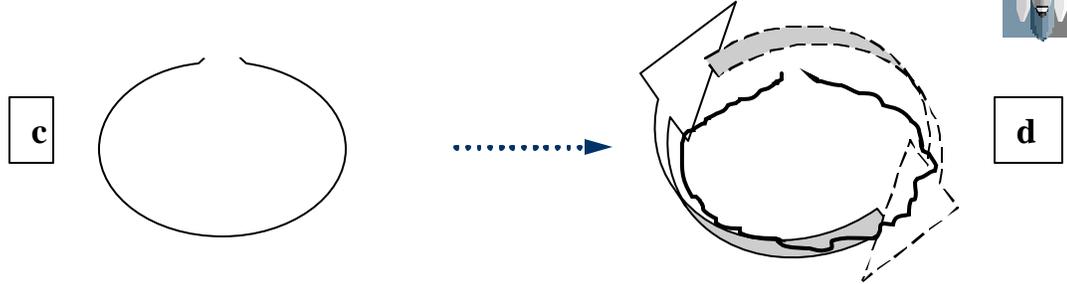

**c**   **d**

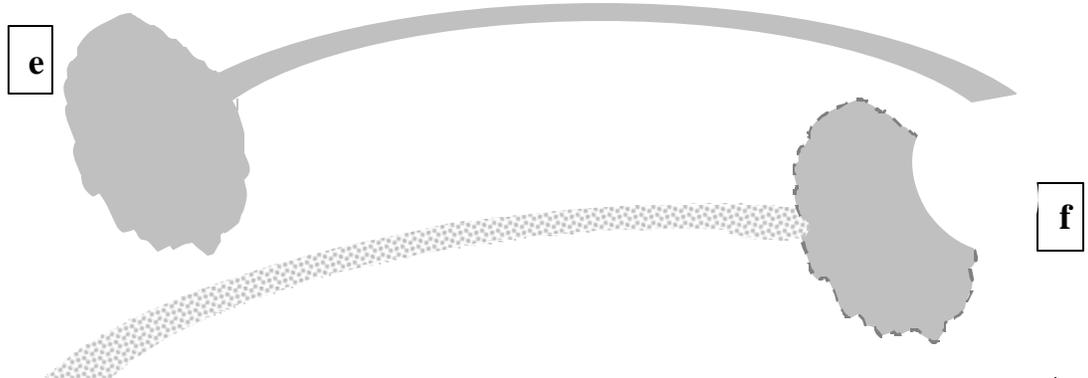

**e**   **f**

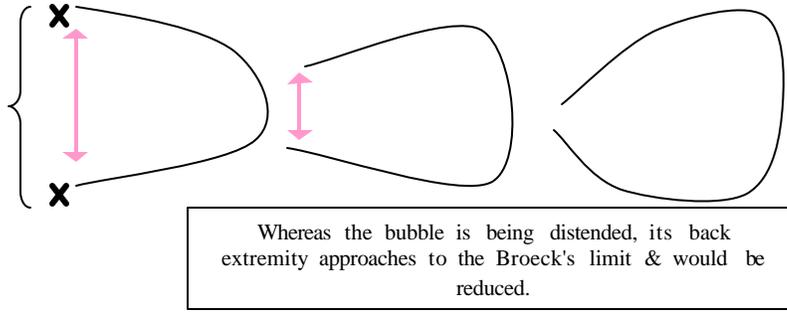

Up & down points of the first stage of a VPW; both need high energies.

Whereas the bubble is being distended, its back extremity approaches to the Broeck's limit & would be reduced.

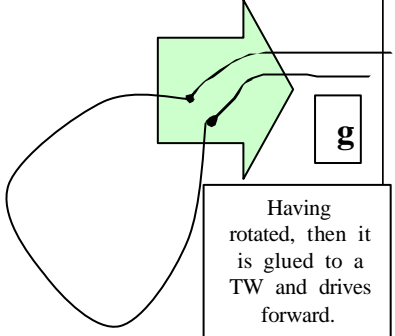

**g**

Having rotated, then it is glued to a TW and drives forward.

***Box of figs* 7**: Two VPWs or more exactly, "system of Alcubierre WD, attached to Van Den Broeck trick, subjected to Krasnikov-like modifications, glued through a surgery to a dynamic asymmetric TW, which has intelligent demonstrations". However, note an embedding diagram for such a dynamic spacetime is misleading, but as a naive approximation of what one can visualize, that works.

Now a proper scenario which seems perfect in the most general condition: Combine the best TWs metric to the best subluminal Alcubierre-Broeck bubble, both asymmetric and dynamic. As a surprise, in the mentioned context, due to presence of WD, the drawback of bad tidal forces – at least inside the bubble – is solved.

At first / (last) stage, a macroscopic hole for entrance / (egress) of the passengers is created (also see Box **8**). Thus in a safe tidal process, they go into a bubble with a possibly least amount of outer surface. They stay therein until two factors take them in the last stage. First operator is the motion of spacetime as the common WD, i.e., expansions & contraction mechanism of the bubble on *not* flat spacetime background but nontrivial topology & geometry of a TW; further, note to the difference of "closed" bubble in WD to the case of VPW. It is open at least in one (junction to the throat) point. Second operator is the reduction of the distance being passed by TW geometry. Obviously, more speed of WD, more need to EM, and less need to TW for original purpose. More efficiency of TW, less need to speed of WD and less need to EM to support the WD, but more need to EM because dynamic properties of TW must now compensate the duty of WD (carrying the bubble) and since dynamic factor has a deep dependence to EM, a long-lived dynamic process needs more EM for more range. Therefore, an elaborate balance between efficiency of WD element and of TW element will render an overall efficiency of the VPW system [126,127]. Explicitly, the directions can be as follows:

**1:** (**7c**) Make a bubble by low NE [88]; **2:** Turn it dynamic & asymmetric (**7d**); needed NE can be much reduced; **3:** Now do a surgery; match the bubble to a TW (**7e** or **7g**).

The main effort is for the entrance mouth, elsewhere mainly need precision. After the VPW would carry the travelers near the destination, two possibilities arise. One is expanding the junction point to come out of the bubble, while another is the gradual disappearing of the egress bubble by spreading the PED (**7f**); although that's technically simpler, but within such dynamic conditions, one should note to balance among low speed of bubble, its amount of expansion, the thickness of bubble (seemingly makes a quantum gravity analysis unavoidable), and a bearable mechanism in respect to tidal forces.

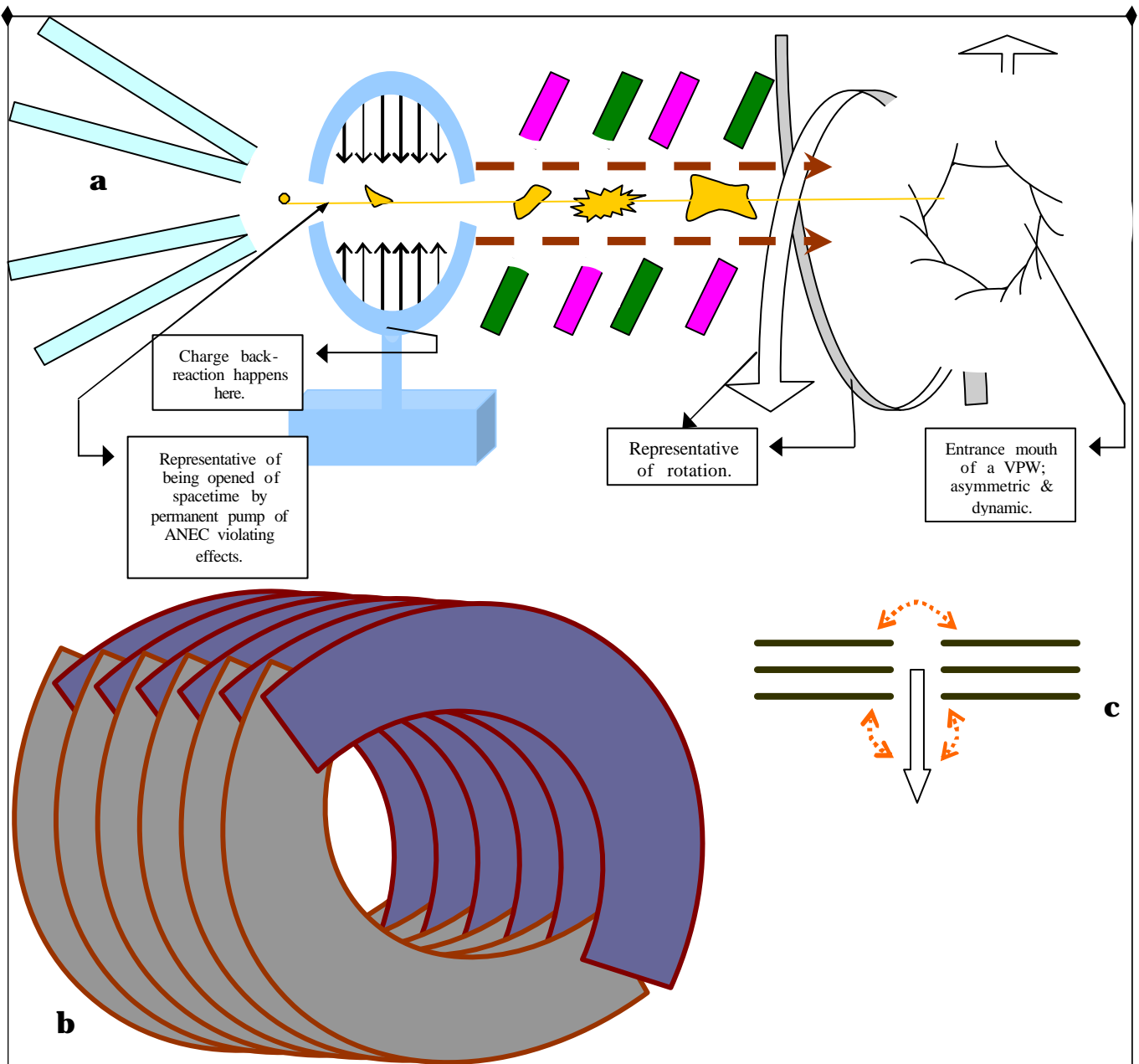

***Box of figs 8***: A design to verify the Eq (7) in the lab which creates a macroscopic mouth and its microscopic controllable continuation. As another application, by suitable manipulations to gain a stream of NE, one can consider such a configuration as the core of a EM maker machine [see **§** below].

Fig **8a**, shows after summation all arts of ANEC violating facilities, – denoted by long light turquoise rectangles & described in Boxes **1**,**2**,**3** – a macroscopic hole is produced; This stage is a difficult beginning (due to initial high curvatures, but fortunately the scales are not big). Actually, that is launched with a tiny hole in spacetime by a pulse of ANEC violation as in figs **1c**, **1g**, **2a**, **2b**, **2g**, **3f**. In the next stage, a sudden powerful Coulomb force may give great contributions; in which the resulted hole in spacetime gets subjected to an electric back-reaction (shown by against vectors). Then, AWEC and/or ANEC violating sets (along with their WEC/NEC violating by-products) come to 'train' the initial configuration, where green (pink) cases denote finalist tails of AWEC (ANEC) violating producers. The distances between the colorful boxes correspond (are proportional) to domain & range of definition of *l* and *m*, subjected to related cut-offs. The ANEC ones need more elegancy because in a *rigorous* model, *all* the time, ANEC should be violated to keep the usable vacuum bulk opened (shown by golden line). The final output is a macroscopic mouth, ready to accept a passer. The curved arrows denote torsion in the metric at last stage (or previous stages).

In (**8b**), the shells are quite movable according to give various arrangements and are contained of the NE leakages publisher rings (an assemble of **8c**). Probably (**8a**) is a pre-launch for second (up to n-th time) step to reach to a VPW.

Fig **8c**, indicates the result may be gained by numerous collections of last equipments in fig **8a**, like the styles in figs **2a**,**2b**,**2c**,**2e**,**2g**.

There are two methods for start; **I:** The passengers place in a definite location, then the entrance mouth comes probing & swallow them forward, **II:** They are placed in a close distance of last out-part of the figs (**8a** or **8c**) and the macroscopic entrance mouth (i.e., between them) to be entered inward; then immediately after entering to the bubble, the traveler sees behind mouth would be diminished severely (e.g., $10^{-10}$ m) and would be attached to throat of a TW, to slide. Hopefully, most difficulties are belong to some initial moments – mouth being held open for travelers to go into the bubble – thereafter, the situation gets much simpler for engineers (due to microscopic scales).

Indeed, computers should yield the best arrangements, and the security margins to not annihilate, if a concentrated positive mass enters to the system to convert such a dynamic VPW, e.g., to numerous tiny evaporating black holes [47,50].

**§**: Exchange of matter between both wormhole mouths can modify their mass ratio starting a process that could lead to a large (stellar-size) negative mass in one of the mouths [see 3, 108, 109].

## Outlook

Let us evaluate current conditions. Based on a paper of Fewster and Roman [102], (after a primary ANEC violation) a "static spherically symmetric QIs satisfying" TW of Kuhfittig [82], requires the radial proper distance of $\ell \geq 62137$ Miles, and NEC violation lied on points at radius $r = 2r_0 - r_2$, to be able to pass a humanoid traveler (along with further possible modifications to cut off the EM region up to a reasonable model).

Alright; but what can we concluded? Indeed, by assuming the [82] as the seminal framework, some corrections were imposed. Dependence to 'time' through two factors of $\Omega$ & $\Xi$, along with shape and redshift functions, both time-dependent should be considered as the first step to improve the scenario in [82]. Second correction was the proposal of removing the 'spherical symmetry' (ideally, any type of symmetry), cast into the discussion on the 'axial symmetry'. Third correction was the encouraging scheme of circumventing the QIs by physical models of CEC violation by interacting fields [6,22], and spacetime metrics [88], in addition of engineering tricks (see Box **3**, also note the detailed arrangements e.g., as in fig **2b**, may cause the equipments to find huge dimensions, perhaps as big as an accelerator). Fourth correction was pointing out the advantages of combination of a TW to a WD; in that case one can have much better view to EM needed for the TW element, through reducing the throat size to about $10^{-10}$ m (that can be more or less, because the limitations come from control engineering); and more pleasant feeling to tidal forces through traveling inside a roomy bubble. The vital concentration to that end is the establishment of a constructive balance to various feature of the elements. Eventually, taking into account the role of charge backreaction to the configuration was other correction which attracts special hopes.

The abstract of above considerations is observed in Eq (7). Although, there is quite possible to impose more complexities to its components, but in the spirit of all the paper, verifying various sample functions of *l* & *m* by a computer seems an approach better than theoretical studies (i.e., experimental tests).

Besides, research on novel material structures is the other branch of breakthrough to the mentioned purposes. For instance, after trivial importance of composites suitable to detect the NED manifestations, Box **1**, fig **3e**, Eq (3b), on conduction of electricity, interesting superconductors like *long* carbonic nanotubes are appropriate candidates [110]. Another useful substance may be called the new stretchable gold conductors [111]. The *local* antigravity effects [114], originated by the spacecraft traveling through the TW can shield the passengers (and the spaceship itself) of possible huge tidal forces. The characteristic of locality is for the least interference of the system and saving energy. One scenario is introduced in [115]. Actually, more solving a problem by different mechanisms, more freedom of other parameters.

Simultaneous to any sort of failure, there would result no NE supporting and the spacetime would become flat. However, dependent to points of passage, the security is rather high. It means if due to any problem the system fails, the spacetime would become flat & the traveler might be appeared e.g., in a jungle! Other cases are: severe tidal constraints, dangerous electricity and being careless of humanoid passengers in e.g., going out of the bearer spacecraft.

Besides, since the r.h.s. of (A1) is to be supported (mostly) by quantum effects (via Casimir energy), some features in macroscopic scales are expected; e.g., the throats might be quantized [81,116]. Also, one possibility of giving contribution to rotation & charge as carriers of exoticity, can be inspired from 'teleforce' discussion [117].

At last, as the first generation of practical spacewarps, by a somehow $\ell$ reduction, one might not serve to passage of humanoid travelers, but traversal of some strategic items like gas, oil, oxygen, or minerals – because of less traversability implications – would give great motivations to improve the field to 'any' kind of application.

order of $c$, the only possibility to dynamic Casimir effect, is to accumulate the effect gradually under the resonance conditions. Also, considering Figs (1 & 2) of [6], vertical motion is another factor (an amplitude nearly equal to hole's radius seems sufficient). Therefore along with being careful, one shouldn't be discouraged according to arguments like p. 17 of [29]. Another good news is, as discussed in division VII of [6], shifting the boundary conditions can have desired influences on ANEC (a flash at least); however, (un?)fortunately, as Figs (3 & 4) – right ones – the common negative values are about 1 (both functions go beneath the *z* axis), that is just for domain wall model & is not seen for perfect mirrors (dotted lines in Fig 4).

[60] If the creation of repulsive mouth be possible, by once trapping of a negative pulse, the process would find its way, e.g., having a NE in a (closed) spiral procedure, the extraction of macroscopic amounts of EM – perhaps after thousands of rotations like in centrifuges – is closer to reality. For example, the different trajectories might give important contributions of the sifting. In that view, the initial ANEC violation pulse is similar to a slow neutron, prerequisite for a big chain reaction in nuclear physics.

[61] T. A. Roman, "Inflating Lorentzian wormholes", Phys. Rev. D47 (1993) 1370, gr-qc/9211012

[62] M. Morris, K. Thorne, "Wormholes in spacetime and their use for interstellar travel: a tool for teaching general relativity" Am. J. Phys. 56, 395 (1988).

[63] D. Hochberg, M. Visser, "Dynamic wormholes, anti-trapped surfaces, and energy conditions", Phys. Rev. D58 (1998) 044021, gr-qc/9802046

[64] D. Hochberg, M. Visser, "General Dynamic Wormholes and Violation of the Null Energy Condition", gr-qc/9901020

[65] E. Poisson, M. Visser, "Thin-shell wormholes: Linearization stability", Phys. Rev. D52 (1995) 7318, gr-qc/9506083

[66] Within dynamical TWs, it has been shown that multiplying a static spherically line element by an overall time dependent conformal factor it is possible (depending on the explicit form chosen for the conformal factor) to either postpone the violation of WEC, to relegate it to the past, or else to restrict its violation to short intervals of time [67,68]. In fact by suitably adjusting parameters one can make the timespan over which WEC is satisfied as large as one wants. Actually either the TW has a vanishing throat radius which is tantamount to its not being a TW at all or the WEC is violated somewhere at least. However, if the TW were expanding fast enough, it appears possible to avoid any violation of the CECs [9]. But as explained in [67], this is because any observer traveling through the TW would see its 'radius' increasing all the way, & thus it would not qualify as a TW in the usual sense. In other words, to get around the CEC violations, it is indeed possible to temporarily suspend the violations, but only at the heavy expense of totally destroying the flare-out properties of the throat [63], in TW engineering, a few key such effects would be enough; particularly, combination of dance-like moving of travelers to various cut-offs of time to relax other parameters when one meets essentially slowly flaring out TW metrics [82]. Also a special choice of dynamic factor exponentially decreases WEC violation & increase proper radial distances too, unfortunately [61]. Therefore, a naive trust to dynamic properties would be wrong & one should adjust dynamic parameters so that the reasonable traversability & usability conditions be regarded.

Besides, according to "quantum interest", tricks of Box 3 cause warming the tools up; therefore, another advantage of dynamic models is trying to minimize warming the system by compensating following PE waves.

[67] S. Kar, D. Sahdev, "Evolving Lorentzian Wormholes", Phys. Rev. D53 (1996) 722, gr-qc/9506094

[68] L. A. Anchordoqui, S. P. Bergliaffa, D. F. Torres, M. L. Trobo, "Evolving wormhole geometries", Phys. Rev. D57 (1998) 829, gr-qc/9710026

[69] If we would be able to specify the exact properties of pattern of behavior of the TW geometry & related consequences, e.g., by being influenced of electromagnetic field in that spacetime [56], that would then be a great breakthrough in coding of information. For instance, a TW can be so short-lived which only photons can pass through it. By means of many shifts & other changes the output signal can be simulated, something similar to "quantum teleportation". Therefore, the tiniest kinds of (controllable) TWs – of lifetime, dimensions & energy – might be useful in that field.

[70] Particle creation which is a known phenomenon in such circumstances, can be against NEDs & weakens stability. However as a strengthening contribution, the Hawking effect [96] in those conditions might yield the particles with NE [71],

although according to "small total amount of NE" [72,73] involved in present consideration, there certainly are nice differences to black holes physics.
[71] Also, it's been interpreted [9], stationary, axially symmetric models, characterize rotating TWs, so note for fast rotations, $g_{tt}$ becomes positive in some region outside the throat, indicating the presence of an ergoregion where particles can no longer remain stationary with respect to infinity, & one can imagine an infalling particle breaking up into two inside the ergoregion; so it is possible to arrange this breakup so that one of the resulting particles has negative total energy.
[72] M. Visser, S. Kar, N. Dadhich, "Traversable wormholes with arbitrarily small energy condition violations", Phys. Rev. Lett. 90, 201102 (2003), gr-qc/0301003.
[73] P. K. F. Kuhfittig, "Can a wormhole supported by only small amounts of exotic matter really be traversable?", Phys. Rev. D68, 067502 (2003), gr-qc/0401048
[74] As an effort to restrict the violations of CECs to an infinitesimally small thin shell, Visser [75,76] has shown that for non-spherically symmetric, static TWs one can have many null geodesics along which the ANEC is satisfied. In fact, moving the EM around in space causes some observers falling through the TW would not encounter it (a nice property for passing through the TW). This was demonstrated by cutting out holes in Minkowski space, & joining them up with a thin (delta-function) layer of matter. Therefore, such TWs have at least three advantages, along with a challenge in density factor: Better distribution of matter-energy, Easier construction through try for a sphere with small principal curvatures [77]; & Being consistent with the restrictions encountered in detection of NED regions by *some limited* rays as in Boxes **1**,**2**. Further, as in [77], for static, spherically symmetric TWs, the tension minus energy density is just $k_g/(2\pi r)$. Thus the required NED is smaller for large TWs with small surface gravity, both of which are practical requirements for a comfortable TW. The corresponding NE, integrated over a shell of width $\ell$ around the TW, is of the order of $2\ell r k_g$, so increased with TW size, but is still smaller for small surface gravity.

Indeed, elaborate geometrical & topological manipulations are so important. For instance, a topological gymnastics in improving energy conditions in WD metric [78] can deeply decrease the needed NE. See [79,86,87,88].
[75] M. Visser, "Traversable wormholes from surgically modified Schwarzschild spacetimes", Nuclear Physics B328, 203 (1989).
[76] M. Visser, "Traversable wormholes: Some simple examples", Phys. Rev D39, 3182 (1989).
[77] D. Ida, S. A. Hayward, "How much negative energy does a wormhole need?", Phys. Lett. A260 (1999) 175, gr-qc/9905033
[78] M. Alcubierre, "The warp drive: hyper-fast travel within general relativity", Class. Quant. Grav. 11 (1994) L73, gr-qc/0009013
[79] http://www.npl.washington.edu/AV/altvw99.html
[80] C. J. Fewster, T. A. Roman, "On wormholes with arbitrarily small quantities of exotic matter ", Phys. Rev. D72 (2005) 044023, gr-qc/0507013
[81] S. W. Kim, H. Lee, "Exact solutions of charged wormholes", Phys. Rev. D63 (2001) 064014, gr-qc/0102077.
[82] P. K. F. Kuhfittig, "Wormholes supported by small amounts of exotic matter: some corrections", gr-qc/0508060
[83] A. Carlini, V. P. Frolov, M. B. Mensky, I. D. Novikov, H. H. Soleng, gr-qc/9506087
[84] G. W. Gibbons, S. W. Hawking, *Commun. Math. Phys.* 148, 345 (1992)
[85] M. Visser, "Traversable wormholes: the Roman ring", Phys. Rev. D55 (1997) 521 gr-qc/9702043
[86] C. Van Den Broeck, "A 'warp drive' with more reasonable total energy requirements", Class. Quant. Grav. 16 (1999) 3973, gr-qc/9905084

everywhere is not a good vacuum!). Also, various cut-offs are important to a high security to that direction. Obviously, more capability of producing NED, more reduction of such obstacles. I would like to thank Fernando Loup who caused me to add these points.

[127] Since change of topology is extremely more effective to reach to remote stars than any kind of not-topology-changing propulsions (e.g., common WD on 4D a Minkowski background), one has to deal with TW element more than WD, but there are other features for a VPW in practice. For example, it's better to be traversed inside the WD element fast to not meet spatial dust and stones. Although, TW element is going to be held fine-sized but the probability of bad perturbations on the system – except for checked, clean paths; probably not-interstellar! – should be verified in the lab.

URL: http://www.mansouryar.com,           Email: mmwormhole@yahoo.com

## *Appendix*

Actually, the nature is excessively conservative to gravitationally change the fabric of spacetime. In a mathematical level, that can be seen in extreme coefficients (huge or infinitesimal) in any theory of gravitation. The situation gets worse when one thinks about "experiments" (generally, not in cosmic scales) in semi-classical GR, because of tiny demonstrations of quantum effects (i.e., magnitudes proportional to $\hbar$). Therefore, an approach might be splitting the parameters into many terms, leading to a weaker distribution of the extreme constants which are arisen in any attempt of (macroscopic) influence on the structure of the known universe.

For example, in spherically symmetric TWs, just 3 terms of stress-energy tensor are calculated and other terms are vanished by the symmetry.

Indeed, the constants of the r.h.s. of the Einstein equations

$$G_{mn} = (8\pi G/c^2) T_{mn} \qquad (A1)$$

impose their presence in a severe manner, but what would happen if one expands the $T_{mn}$ to more terms in the asymmetric geometries? Principally, in a mathematical point of view, than can reduce the power of extreme constants. Let us review the possibilities of the generalizations the initial scenario of MT which are in that direction.

In [118] authors state: In any static spacetime one can decompose the spacetime metric into block diagonal form:

$$ds^2 = g_{mn} dx^m dx^n = -\exp(2f)dt^2 + g_{ij} dx^i dx^j \qquad (A2)$$

Also they give a very general definition of MT flare-out condition:

$$\exists l_*^-, l_*^+ > 0: \qquad \forall l \in (-l_*^-, 0) \cup (0, l_*^+) \qquad G_{\hat{t}\hat{t}} + G_{\hat{r}\hat{r}} < 0 \qquad (A3)$$

where in terms of the total stress energy reads:

$$\exists l_*^-, l_*^+ > 0: \qquad \forall l \in \left(-l_*^-, 0\right) \cup \left(0, l_*^+\right) \qquad T_{\hat{t}\hat{t}}^{total} + T_{\hat{r}\hat{r}}^{total} < 0 \qquad (A4)$$

such consideration causes its own consequences (about geometry, CECs, properties of the throat – which is defined as a 2D hypersurface of minimal area taken in one of the constant-time spatial slices –, flaring out conditions, gravitational potential, transverse pressures, etc) [8]. Partially, they expand their results to dynamic cases [63, see Table 1 therein, 64, 119].

The dynamic geometries encounter interesting phenomena like having two throats (being in casual contact to be called traversable), but there are limitations too; e.g., if we identify the torsion with that appearing naturally in the spectrum of closed strings, then we find it actually worsens the violations of the NEC at the throats [63].

On the other hand, Hochberg & Visser consider the

$$ds^2 = -e^{2y}\left(1 - \tfrac{2m}{r}\right)dv^2 + 2e^y \, dv dr + r^2\left(d\boldsymbol{q}^2 + \sin^2\boldsymbol{q} \, d\boldsymbol{f}^2\right) \qquad (A5)$$

as the most general metric describing a time-dependent spherically symmetric spacetime, adaptable to describe an inter-universe TW.

In [120], authors declare: The metric for any static and spherically symmetric spacetime which might contain a throat, i.e., $r(0) > 0$, can be cast into the form

$$ds^2 = -f(l)dt^2 + dl^2 + r^2(l)\left(d\boldsymbol{q}^2 + \sin^2\boldsymbol{q} \, d\boldsymbol{f}^2\right) \qquad (A6)$$

In this metric, the semiclassical (A1) takes the form

$$G_{\boldsymbol{m}}^n(f(l), r(l)) = 8\boldsymbol{p}\left\langle T_{\boldsymbol{m}}^n(f(l), r(l); \boldsymbol{x})\right\rangle \qquad (A7)$$

where $\boldsymbol{x}$ is the non-minimal scalar coupling to the metric. Then, they represent some calculations.

In [9], Teo discusses a stationary, axially symmetric TW, starting by

$$ds^2 = g_{tt}dt^2 + 2g_{t\boldsymbol{j}}dt d\boldsymbol{j} + g_{\boldsymbol{jj}}d\boldsymbol{j}^2 + g_{ij}dx^i dx^j \qquad (A8)$$

and promotes the MT metric to a more general condition. It has some advantages to CECs [105], in addition of some new problems; e.g., one should consider the possible presence of an ergoregion surrounding the throat or perhaps more corrections would be essential [121].

Other similar version, is the work of Khatsymovsky [107]. He chooses the metric

$$ds^2 = \exp(2\Phi)dt^2 - d\boldsymbol{r}^2 - r^2[d\boldsymbol{q}^2 + \sin^2\boldsymbol{q}(d\boldsymbol{f}^2 + 2hd\boldsymbol{f} \, dt)] \qquad (A9)$$

where $h(\boldsymbol{r}, \boldsymbol{q})$ has the sense of angular velocity of the local inertial frame and will be called the angular velocity of rotation.

In [35] authors start by a parameterized surface of revolution as

$$x(r, \boldsymbol{f}) \coloneqq \left(x^1(r, \boldsymbol{f}), x^2(r, \boldsymbol{f}), x^3(r, \boldsymbol{f})\right) = (r\cos(\boldsymbol{f}), r\sin(\boldsymbol{f}), P(r)) \qquad (A10)$$

where $r = \sqrt{(x^1)^2 + (x^2)^2}$ and $f$ is shown in fig 2 therein. Then, the induced metric on the surface is given by $ds^2 = [1 + P_{,1}(1)^2]dr^2 + r^2 df^2$ which yields:

$$ds^2 = -e^{g(r)}dt^2 + [1 + P_{,1}(1)^2]dr^2 + r^2\left(dq^2 + \sin^2 q \, df^2\right) \tag{A11}$$

Besides, the matter field is assumed an anisotropic fluid with the stress energy tensor of:

$$T_{mn} = (r + p_\perp)u_m u_n + p_\perp g_{mn} + (p_q - p_\perp)s_m s_n \quad , \quad u^m u_m = -1, \quad s^m s_m = 1, \quad u^m s_m = 0 \tag{A12}$$

along with the conservation law
$$T^{mn}{}_{;n} = 0 \tag{A13}$$

Another example is a metric reported by Krasnikov [122]:

$$ds^2 = -dt^2 + 2t/r \, dt dr + [1 - (t/r)^2]dr^2 + (BK_0 r)^2\left(dq^2 + \sin^2 q \, df^2\right) \tag{A14}$$

with a claim of solving the WEC violation.

In that spirit, one can refer to [90,105]. In [105], adding the rotation to a static spherically symmetric model, severely decreases the magnitude of WEC violation.

However, there are some correlated affairs in that nonlinear scenario. First one is inventing suitable metrics of TWs and the second one is the needed stress-energy tensors to that metrics. The important thing is review the constraints imposed by the nature to that desire. Obviously, technological limits have most major effects on $T_{mn}$. In that direction, since $T^{ij}$ is the $x^i$ component of momentum per area in the $x^j$ direction (describing a shearing from stresses), much engineering is required as introduced in Box **2**. The other factor can be considered as the finite options to geometry due to the effect of stress-energy tensor to l.h.s. of (A1) [58].

Generally, we have [3]:

$$T_{mn} \equiv \begin{bmatrix} r & \vdots & S_j \\ \cdots & \cdot & \cdots \\ S_i & \vdots & p_{ij} \end{bmatrix} \tag{A15}$$

where $r$ is the energy density, $S_i$ is the energy flux (essentially a generalization of the Poynting vector), and $p_{ij}$ is the stress (essentially a generalization of the notion of pressure). Therefore, the strategy should be extraction of all the abilities of the matrix elements. For instance, one might meet

$$\sim 10^{40} \times T_{mn} = \begin{bmatrix} T_{tt} & T_{tr} & T_{tq} & T_{tf} \\ T_{rt} & T_{rr} & T_{rq} & T_{rf} \\ T_{qt} & T_{qr} & T_{qq} & T_{qf} \\ T_{ft} & T_{fr} & T_{fq} & T_{ff} \end{bmatrix} \tag{A16}$$

in the case of spherical coordinates; and deduces the power of constant coefficient divided by the number of matrix elements; $40/16 = 2.5$ as the share of every element.

Having in mind the various technical cut-offs (Box **4**) within the program of stress-energy tensor extension, reducing the quantity of magnitude of order, can be called the "puffing" contributions, where accelerate a flare-out, whereas one has already accepted a slow flaring out Kuhfittig model [82] as the base metric.

Thus, the consequences can be seen to other relations e.g., Eq (B3, B4), whereas adding to metric would complicate the terms of those Eqs than just a radial component to a multi-dimensional 'dance' of travelers. However, focusing on other geometrical adjustments [88, 106] have their own results.

Also, note to the time dependent redshift, e.g., $\boldsymbol{g}(r) = -\boldsymbol{l}(r,t)$, then WEC is $\boldsymbol{r}(r)c^2 - \boldsymbol{t}(r,t) \geq 0$, as considered in [90]:

$$\boldsymbol{r}(r)c^2 - \boldsymbol{t}(r,t) = \frac{1}{8\boldsymbol{p}Gc^{-4}} \left[ \frac{2}{r} e^{-2\boldsymbol{a}(r-r_0)} \left( \boldsymbol{a}'(r - r_0) - \frac{\partial}{\partial r} \boldsymbol{l}(r,t) \right) \right] \quad (A17)$$

Effect on all components can distribute the $c^4/8\boldsymbol{p}G$; that's the same in the case of more role for time:

$$\underbrace{\boldsymbol{r}c^2 - \boldsymbol{t} \pm 2\boldsymbol{f}}_{\textit{This side must be made longer}} = \frac{1}{8\boldsymbol{p}Gc^{-4}} \left[ \frac{2}{r} e^{-2\boldsymbol{a}(r,t)} \left( \frac{\partial}{\partial r} \boldsymbol{a}(r,t) - \frac{\partial}{\partial r} \boldsymbol{l}(r,t) \right) \pm \frac{4}{r} e^{\boldsymbol{l}(r,t)} e^{-\boldsymbol{a}(r,t)} \frac{\partial}{\partial t} \boldsymbol{a}(r,t) \right] \quad (A18)$$

or taking into account of rotation (below equation,) [105], combination of both of them, etc.

$$\boldsymbol{r} - \boldsymbol{t} = re^{-2\boldsymbol{m}} \left( \frac{\partial \boldsymbol{l}}{\partial r} + \frac{\partial \boldsymbol{l}}{\partial r} \sin^2 \boldsymbol{q} \right) + re^{-2\boldsymbol{m}} \left( \frac{\partial \boldsymbol{m}}{\partial r} + \frac{\partial \boldsymbol{m}}{\partial r} \sin^2 \boldsymbol{q} \right) + \left( -\frac{1}{2} \right) \boldsymbol{w} \ r^3 \frac{\partial \boldsymbol{w}}{\partial r} e^{-2\boldsymbol{l}} e^{-2\boldsymbol{m}} \sin^2 \boldsymbol{q} \quad (A19)$$

*More and more terms*

Based on that strategy, – in addition of $T_{\hat{t}\hat{t}} = \boldsymbol{r}$ energy density, $T_{\hat{r}\hat{r}} = \boldsymbol{t}$ radial tension pressure per unit area, and $T_{\hat{q}\hat{q}} = T_{\hat{f}\hat{f}} = \boldsymbol{p}$ lateral pressure, in spherically symmetric case – one can interpret $T_{\hat{t}\hat{r}} = \pm \boldsymbol{f}$ as energy flux [61,90], $T_{\hat{t}\hat{f}}$ as the rotation of the matter distribution [105], and more and more different terms (up to 16 ones) with unequal interpretations.

Trivially, correlated to consequences for the geometrical part of the scheme; in the form of longer equations [105]:

$$\left| R_{\hat{2}\hat{0}\hat{2}\hat{0}} \right| = \left| R_{\hat{3}\hat{0}\hat{3}\hat{0}} \right| = \boldsymbol{g}^2 \left| R_{\hat{q}\hat{t}\hat{q}\hat{t}} \right| + \boldsymbol{g}^2 \left( \frac{v}{c} \right)^2 \left| R_{\hat{q}\hat{r}\hat{q}\hat{r}} \right| + 2\boldsymbol{g}^2 \left( \frac{v}{c} \right) \left| R_{\hat{q}\hat{t}\hat{q}\hat{r}} \right|$$

$$= \boldsymbol{g}^2 \left| \frac{1}{r} \frac{\partial}{\partial r} \boldsymbol{l}(r,t) e^{-2\boldsymbol{a}(r,t)} \right| + \boldsymbol{g}^2 \left( \frac{v}{c} \right)^2 \left| \frac{1}{r} \frac{\partial}{\partial r} \boldsymbol{a}(r,t) e^{-2\boldsymbol{a}(r,t)} \right| + 2\boldsymbol{g}^2 \left( \frac{v}{c} \right) \left| \frac{1}{r} \frac{\partial}{\partial t} \boldsymbol{a}(r,t) e^{\boldsymbol{l}(r,t)} e^{-\boldsymbol{a}(r,t)} \right| \quad (A20)$$

Also, procedure of weakening the coefficients should avoid severe fluctuations; i.e., related components of the Riemann curvature tensor should change smoothly, regarding sensitivity of flaring out, characteristics of key intervals and small amount of involved NE.

Theorems stated by some authors [58,107] should be considered in a detailed analysis of arisen constraints. Those can help to study by vanishing various derivatives, etc.

Because of not being any symmetry (to overcome tidal forces), the path of a traveler to pass through a TW is not 'general' at all & quite 'special' instead. That is completely dependent to some things, i.e., direction of approach, velocity, dynamic features like twist, rotation, spin, reasonable sensitivity and locating on particular places depending on relations of variables to each other. Briefly, the program of any travel operation is quite exclusive [123].